\documentclass{aa}  

\usepackage{graphicx}
\usepackage{txfonts}
\usepackage{rotating, graphicx}

\newcommand\Tstrut{\rule{0pt}{2.6ex}}         % = `top' strut
\newcommand\Bstrut{\rule[-1.4ex]{0pt}{0pt}}   % = `bottom' strut

\begin{document}

   \title{Overdensity of submillimeter galaxies around the $z\simeq 2.3$ MAMMOTH-1 nebula}

   \subtitle{The environment and the powering of an Enormous Lyman-Alpha Nebula}

   \author{F. Arrigoni Battaia
          \inst{1}
          \and
          Chian-Chou Chen
	  \inst{1}
	  \and
	  M. Fumagalli
	  \inst{2,3}
	  \and
	  Zheng Cai
	  \inst{4}
	  \and
	  G. Calistro Rivera
	  \inst{5}
	  \and
	  Jiachuan Xu
	  \inst{6}
	  \and
	  I. Smail
	  \inst{3}
	  \and
	  J. X. Prochaska
	  \inst{4}
	  \and
	  Yujin Yang
	  \inst{7}
	  \and
	  C. De Breuck
	  \inst{1}
          }

   \institute{European Southern Observatory, Karl-Schwarzschild-Str. 2, D-85748 Garching bei M\"unchen, Germany\\
              \email{farrigon@eso.org}
         \and
	 Institute for Computational Cosmology, Durham University, South Road, Durham DH1 3LE, UK
	 \and 
	 Centre for Extragalactic Astronomy, Durham University, South Road, Durham DH1 3LE, UK
	 \and
	 UCO/Lick Observatory, University of California, 1156 High Street, Santa Cruz, CA 95064, USA
	 \and
	 Leiden Observatory, Leiden University, P.O. Box 9513, 2300 RA Leiden, The Netherlands
	 \and
	 Tsinghua University, 30 Shuangqing Rd, Haidian Qu, Beijing Shi, China
	 \and
	 Korea Astronomy and Space Science Institute (KASI), 776 Daedeokdae-ro, Yuseong-gu, Daejeon, 34055, Korea
             }

   \date{Received ---; accepted ---}

% \abstract{}{}{}{}{} 
% 5 {} token are mandatory
 
  \abstract{In the hierarchical model of structure formation, giant elliptical galaxies form through merging processes within the highest density peaks known as protoclusters.
   While high-redshift radio galaxies usually pinpoint the location of these environments, we have recently discovered at $z\sim2-3$ three Enormous ($>200$~kpc) 
   Lyman-Alpha Nebulae (ELANe) that host multiple Active Galactic Nuclei and that are surrounded by overdensities of Lyman-$\alpha$ Emitters (LAE). 
   These regions are prime candidates of massive protoclusters in the early stages of assembly. To characterize the star-forming activity within these rare structures -- 
   both on ELAN and protocluster scales -- we have initiated an observational campaign with the James Clerk Maxwell Telescope (JCMT) and the Atacama Pathfinder EXperiment (APEX) telescopes.  
   In this paper we report on sensitive SCUBA-2/JCMT 850 and 450~$\mu$m observations of a $\approx128$~arcmin$^2$ field comprising the ELAN MAMMOTH-1, together with the peak of 
   the hosting BOSS1441 LAE overdensity at $z=2.32$. These observations unveil $4.0\pm1.3$ times higher source counts at 850~$\mu$m with respect to blank fields, likely
   confirming the presence of an overdensity also in obscured tracers. We find a strong detection at 850~$\mu$m associated with the continuum source embedded within the 
   ELAN MAMMOTH-1, which -- together with the available data from the literature -- allow us to constrain the spectral energy distribution of this source to be of an Ultra-Luminous Infrared Galaxy (ULIRG) with a far-infrared
   luminosity of $L_{\rm FIR}^{\rm SF}=2.4^{+7.4}_{-2.1}\times10^{12}$~L$_{\odot}$, % OLD VALUE %% $L_{\rm FIR}=8.9\times10^{12}$~L$_{\odot}$, 
   and hosting an obscured AGN. Such a source is thus able to power the hard photoionization plus outflow scenario depicted in \citet{Cai2016}
   to explain the extended Lyman-$\alpha$, \ion{He}{ii}$\lambda1640$ and \ion{C}{iv}$\lambda1549$ emission, and their kinematics. In addition, the two brightest detections at 850~$\mu$m ($f_{850}>18$~mJy) sit at the density peak of the
   LAEs overdensity, likely pinpointing the core of the protocluster. Future multiwavelength and spectroscopic datasets targeting the full extent of the BOSS1441 overdensity have the
   potential of firmly characterizing a cosmic nursery of giant elliptical galaxies, and ultimately of a massive cluster. 
   }
   %{In the hierarchical model of structure formation, giant elliptical galaxies form through merging processes within the highest density peaks known as protoclusters.
   %While high-redshift radio galaxies usually pinponint the location of these environments, we have recently discovered at $z\sim2-3$ three Enormous Lyman-Alpha Nebulae (ELAN)
   %that host multiple Active Galactic Nuclei and that are surrounded by overdensities of Lyman-Alpha Emitters. 
   %These regions are prime candidates of massive protoclusters in the early stages of assembly.  
   %{bla bla
   %{bla bla
  % { bla bla

   \keywords{Submillimeter: galaxies -- Galaxies: high-redshift -- Galaxies: halos -- Galaxies: clusters: general -- (Cosmology:) large-scale structure of Universe -- Galaxies: evolution
               }

   \maketitle
%
%-------------------------------------------------------------------

\section{Introduction}

In the present-day Universe, giant elliptical galaxies are found at the centers of massive clusters. 
Being characterized by old, and coeval stellar populations, these central galaxies must have 
formed the bulk of their stars in exceptional star-forming events at early epochs, or must have accreted several coeval 
galaxies (e.g., \citealt{Kauffmann96}). Indeed, the current hierarchical structure formation model predicts 
that these central galaxies merge with %predate 
several nearby satellite galaxies to build up their stellar mass (e.g., \citealt{West1994}). This violent merging process is thought to 
take place in the highest density peaks in the early Universe, in the so-called protoclusters.   
Despite that a lot of effort has been put in characterizing overdensities of galaxies at high-redshift, there is still an 
open debate on which is the best technique to map protoclusters and on which systems represent the nurseries of present-day massive clusters, and thus 
the site of formation of elliptical galaxies (e.g., \citealt{steidel00,Venemans2007,Dannerbauer2014,Orsi2016,Cai2017a,Miller2018,Oteo2018}). 

To date, high-redshift radio galaxies (HzRGs) are one of the best candidates for pinpointing the location of these extremely dense environments (\citealt{mileyd08}).
This result is supported by the rarity of these systems, by Ly$\alpha$ emitter (LAEs) overdensities near them, and in some cases by overdensities in 
submillimeter observations (\citealt{Stevens2003,Humphrey2011,Rigby2014,Zeballos2018}). Being the host of an active galactic nucleus (AGN) and characterized by intense radio emission, 
HzRGs are also known for their associated giant Ly$\alpha$ nebulae on hundreds of kpc scales, suggesting the presence of a large 
amount of gas in these systems (e.g., \citealt{rvr+03}). This Ly$\alpha$ emission is a complex result of AGN ionization, jet-ambient gas interaction, 
and intense star formation (\citealt{VillarMartin2003,Vernet2017}).
Despite these pieces of evidence for protoclusters around HzRGs, we have recently discovered enormous Ly$\alpha$ nebulae (ELANe; \citealt{Cai2016}), more extended than those around HzRGs, and in 
even more extreme environments at $z\sim2-3$ (\citealt{hennawi+15,Cai2016,FAB2018}).

The ELANe -- with observed Ly$\alpha$ surface brightness SB$_{\rm Ly\alpha}\gtrsim10^{-17}$~erg~s$^{-1}$~cm$^{-2}$~arcsec$^{-2}$ on $\gtrsim 100$~kpc, maximum extents of 
$>250$~kpc and Ly$\alpha$ luminosities of $L_{\rm Ly\alpha}>10^{44}$~erg~s$^{-1}$ -- represent the extrema of known radio-quiet Ly$\alpha$ nebulosities.
Indeed, previously well studied radio-quiet Ly$\alpha$ nebulae at $z\sim2-6$, {\it a.k.a} Ly$\alpha$ blobs (LABs; e.g., \citealt{steidel00,matsuda04,Yang2010,Matsuda2011,Prescott2015,Geach2016,Umehata2017}), are characterized by smaller luminosities $L_{\rm Ly\alpha}\sim10^{43-44}$~erg~s$^{-1}$, and smaller extents (50-120~kpc) down to similar surface brightness levels (SB$_{\rm Ly\alpha}\sim10^{-18}$~erg~s$^{-1}$~cm$^{-2}$~arcsec$^{-2}$).
While most of the LABs have a powering mechanism that is still debated (e.g., \citealt{Mori2004,Dijkstra2009_hd,Rosdahl12,Overzier2013,fab+15a,Prescott2015b,Geach2016}), ELANe are usually explained by photoionization and/or feedback activity of the associated quasars and companions (\citealt{cantalupo14, hennawi+15, Cai2016, FAB2018}).

The current sample of ELANe still comprises only a handful of 
objects (\citealt{hennawi+15,Cai2016,FAB2018,Cai2018}). All these ELANe are associated with local overdensities of AGN, with up to 4 known quasars sitting at the same redshift of the 
extended Ly$\alpha$ emission for the ELAN Jackpot (\citealt{hennawi+15}). Given the current clustering estimates for AGN,  
the probability of finding a multiple AGN system is very low,  
$\approx 10^{-7}$ for a quadrupole AGN system (\citealt{hennawi+15}).
This occurrence makes a compelling case that these nebulosities are sitting in very dense environments.
This working hypothesis is further strengthen by the detection of a large number of associated LAEs on small (\citealt{FAB2018}) and on large scales (\citealt{hennawi+15, Cai2016}).
Such overdensities of LAEs are comparable or even higher than in the case of HzRGs and LABs (\citealt{hennawi+15,Cai2016,FAB2018}).

Most of the known ELANe (\citealt{cantalupo14,hennawi+15,fab+15b,FAB2018}) show (i) at least one bright type-1 quasar embedded in the extended emission, 
(ii) non-detections in \ion{He}{ii}$\lambda1640$\AA\ 
and \ion{C}{iv}$\lambda1549$\AA\, down to sensitive SB limits ($\sim10^{-18}-10^{-19}$~erg~s$^{-1}$~cm$^{-2}$~arcsec$^{-2}$), and (iii) relatively quiescent kinematics for the
Ly$\alpha$ emission (FWHM$\simeq600$~km~s$^{-1}$) with a single peaked Ly$\alpha$ line down to the current resolution of the instrument used. 

Notwithstanding these results, the ELANe and their environment have been currently studied only in unobscured tracers, possibly resulting in a biased vision of the phenomenon. 
A complete view of these systems requires a multiwavelength dataset. In particular, submillimeter galaxies (SMGs; \citealt{Smail1997}) 
have been shown to be linked to merger events (e.g., \citealt{Engel2010,Ivison2012,Alaghband-Zadeh2012,Fu2013,TC2015,Oteo2016}) and
to be good tracers of protoclusters (e.g., \citealt{Smail2014,Casey2016,Hung2016,Wang2016,Oteo2018,Miller2018}). 
For these reasons, and to directly test whether our newly discovered ELANe could be powered by intense obscured star-formation, 
we have initiated a submillimeter campaign with the James Clerk Maxwell Telescope (JCMT) and the Atacama Pathfinder EXperiment (APEX)  telescopes to map 
the obscured star-forming activity (if any) 
associated with these rare systems and their environment.  

Here we report the results of our observations of the ELAN MAMMOTH-1 at $z=2.319$ (\citealt{Cai2016}) using the 
Submillimetre Common-User Bolometer Array 2 (SCUBA-2; \citealt{Holland2013}) 
on JCMT. This ELAN has been discovered close to the density peak of the large-scale
structure BOSS1441 (\citealt{Cai2017a}). BOSS1441 has been identified thanks to a group of strong IGM Ly$\alpha$ absorption systems (\citealt{Cai2017a}).
Follow-up narrow-band imaging, together with spectroscopic observations have constrained the Lyman-$\alpha$ Emitters (LAEs, i.e. sources with 
rest-frame equivalent width EW$_0^{Ly\alpha}>20$\AA) 
in this field (\citealt{Cai2017a}). With a LAE density of $\approx12\times$ that in random fields in a (15~cMpc)$^3$ volume, 
BOSS1441 is one of the most overdense fields discovered to date.

The ELAN MAMMOTH-1 is unique compared to the other few ELAN so far discovered, showing
(i) only a relatively faint source ($i=24.2$) embedded in it, (ii) extended emission ($\gtrsim30$~kpc) in \ion{He}{ii}$\lambda1640$\AA\ 
and \ion{C}{iv}$\lambda1549$\AA\, and (iii) double-peaked line profiles 
with velocity offsets of $\approx700$~km~s$^{-1}$ for Ly$\alpha$, \ion{He}{ii}, and \ion{C}{iv}. 
In light of these evidences, \citet{Cai2016} explained this ELAN as circumgalactic/intergalactic gas powered 
by photoionization or shocks due to a galactic
outflow, most likely powered by an enshrouded AGN.
With the SCUBA-2 data we can start to better constrain the nature of this powering source.

This work is structured as follows. In Section~\ref{sec:obs}, we describe our observations and data reduction.
In Section~\ref{catalogs} we present the catalogs at 450 and 850~$\mu$m, along with reliability and completeness tests.
In Section~\ref{NC}, we describe how we determined the pure source number counts, estimated the underlying counts model 
through Monte Carlo simulations, and how we used this models to get the true counts. 
The same Monte Carlo simulations allowed us to assess the flux boosting (Section~\ref{fluxBoost}) and the positional uncertainties (Section~\ref{pos_err}) 
inherent to our observations. 
In Section~\ref{sec:results} we show (i) the true number counts and compare them to number counts in blank fields, and 
(ii) the location of the discovered submillimeter sources in comparison to known LAEs. 
We then discuss our overall detections and the counterpart of the ELAN MAMMOTH-1 in Section~\ref{sec:disc}, and we summarize our results in Section~\ref{sec:summ}.

Throughout this paper, we adopt the cosmological parameters $H_0=70$~km~s$^{-1}$~Mpc$^{-1}$, $\Omega_M =0.3$ 
and $\Omega_{\Lambda}=0.7$. In this cosmology, 1\arcsec\ corresponds to about 8.2 physical kpc at $z=2.319$.
All distances reported in this work are proper.

%--------------------------------------------------------------------
\section{Observations and Data Reduction}
\label{sec:obs}
%--------------------------------------------------------------------

The SCUBA-2 observations for the MAMMOTH-1 field were conducted at JCMT during flexible observing in 
2018 January 16, 17, and 18 (program ID: M17BP024) under good weather conditions (band 1 and 2, $\tau_{225{\rm GHz}}\leq 0.07$).
The observations were performed with a Daisy pattern covering $\simeq13.7\arcmin$ in diameter, and were centered at the location of the 
ELAN MAMMOTH-1 as indicated in \citet{Cai2016}. Note however that the exact coordinate of the ELAN MAMMOTH-1 have been refined to be R.A. $=$ 14:41:24.456, 
and Dec. $=$ +40:03:09.45. 
To facilitate the scheduling we divided the observations in 6 scans/cycles of about 30 minutes, for a total of 3 hours.\\

The data reduction follows closely the procedures detailed in \citet{TC2013a}. In short, the data were reduced using the Dynamic Iterative Map Maker (DIMM) 
included in the SMURF package from the STARLINK software (\citealt{Jenness2011,Chapin2013}). The standard configuration file dimmconfig\_blank\_field.lis was 
adopted for our science purposes. Data were reduced for each scan and the MOSAIC\_JCMT\_IMAGES recipe in PICARD, the Pipeline for Combining and Analyzing 
Reduced Data (\citealt{Jenness2008}), was used to coadd the reduced scans into the final maps. 

The final maps were applied standard matched filter to increase the point source detectability, using the PICARD recipe SCUBA2\_MATCHED\_FILTER. Standard flux 
conversion factors (FCFs; 491 Jy pW$^{-1}$ for 450~$\mu$m and 537 Jy pW$^{-1}$ for 850~$\mu$m) with 10\% upward corrections were adopted for flux calibration. 
The relative calibration accuracy is shown to be stable and good to 10\% at 450~$\mu$m and 5\% at 850~$\mu$m (\citealt{Dempsey2013}).

\begin{figure}
\centering
\includegraphics[width=1.0\columnwidth]{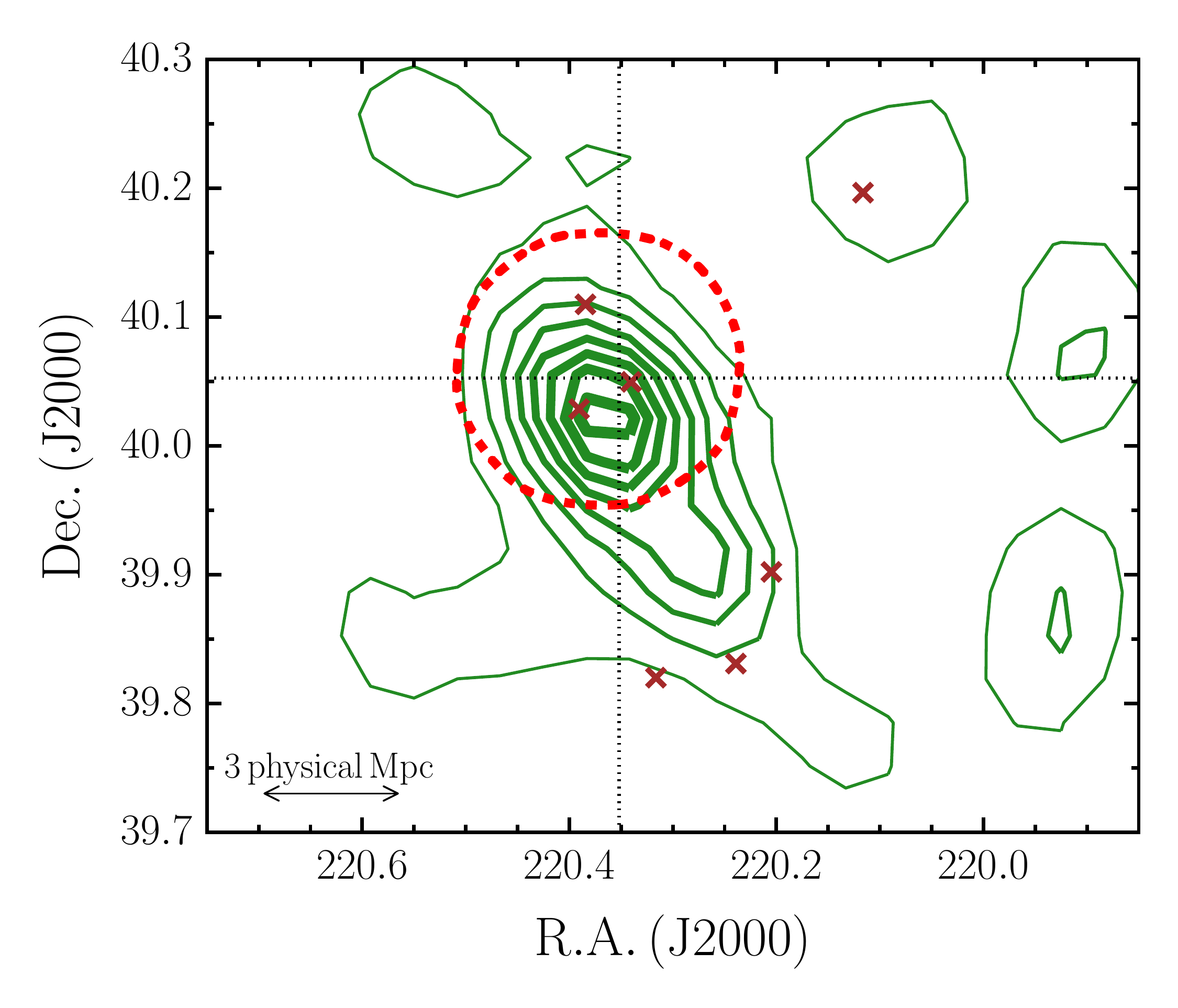}
\caption{The galaxy overdensity BOSS1441 at $z=2.32\pm0.02$ (\citealt{Cai2017a}). The density contours (green) for LAEs are shown in steps of 0.1 galaxies per arcmin$^2$, with the
inner density peak of 1.0 per arcmin$^2$. The density contours are shown with increasing thickness for increasing galaxy number density. The brown crosses indicate the
positions of known QSOs in the redshift range $2.30\leq z < 2.34$, and thus likely within the overdensity. We also highlight the position of the ELAN MAMMOTH-1 (dotted crosshair), 
and the effective area of our SCUBA-2 observations (red dashed contour).} 
\label{BOSS1441}
\end{figure}

The final central noise level for our data is 0.88~mJy/beam and 5.4~mJy/beam, respectively at 850~$\mu$m and 450~$\mu$m.
In the reminder of this work we focused on the regions of the data characterized by a noise level less than three times the central noise.
We refer to this area as effective area.
In Figure~\ref{BOSS1441} we overlay the field-of-view (corresponding to the effective area) of our {SCUBA-2} observations 
(dashed red) on the overdensity of LAEs known from the work of \citet{Cai2017a} (green contours).
In Table~\ref{obs} we summarize the center, the effective area, and the central noise ($\sigma_{\rm CN}$) of our observations.

\begin{table}
\begin{center}
\caption{SCUBA-2 observations around the MAMMOTH-1 nebula.}
\begin{tabular}{lr}
\hline
\hline
 R.A. (J2000; h:m:s)                        		&   14:41:27.62  \\ %\Tstrut\\
 DEC. (J2000; $^\circ$:$'$:$''$)                        &  +40:03:31.4   \\
 Effective Area (850\,$\mu$m; arcmin$^2$)$^{a}$                   &    127.812     \\
 Effective Area (450\,$\mu$m; arcmin$^2$)$^{a}$                   &    126.648     \\
 Central Noise, $\sigma_{\rm CN}$ (850\,$\mu$m; 1\,$\sigma$; mJy/beam)     &       0.88     \\
 Central Noise, $\sigma_{\rm CN}$ (450\,$\mu$m; 1\,$\sigma$; mJy/beam)     &       5.4     \\ %\Bstrut\\
\hline 
\hline 
 \multicolumn{2}{l}{$^{a}$ {\footnotesize Total area to 3 times the central
 noise level.}}
 \Tstrut\\ 
\end{tabular}
\label{obs}
\end{center}
\end{table}

%--------------------------------------------------------------------
\section{Source Extraction and Catalogues}
\label{catalogs}
%--------------------------------------------------------------------

To extract the detections from both maps, we proceed following \citet{TC2013a}.
We first extracted sources with a peak S/N~$>2$ within the effective area of our observations (see Table~\ref{obs}).
Specifically, our algorithm for source extraction finds the maximum pixel within the 
selected region, takes the position and the information of the peak, and subtracts
a scaled PSF centered at such position\footnote{As PSFs for our observations, we adopt the PSFs at 850 and 450~$\mu$m generated by \citet{TC2013b} (see their Figure~2).}. 
The process has been iterated until the peak S/N
went below 2.0. 
These sources constituted the preliminary catalogs at 850 and 450~$\mu$m.
We then cross-checked the two catalogs to find counterparts in the other band.
We considered a source as a counterpart if its position at 450~$\mu$m lays within the
850~$\mu$m beam.

\begin{figure*}
\centering
\includegraphics[width=18.5cm]{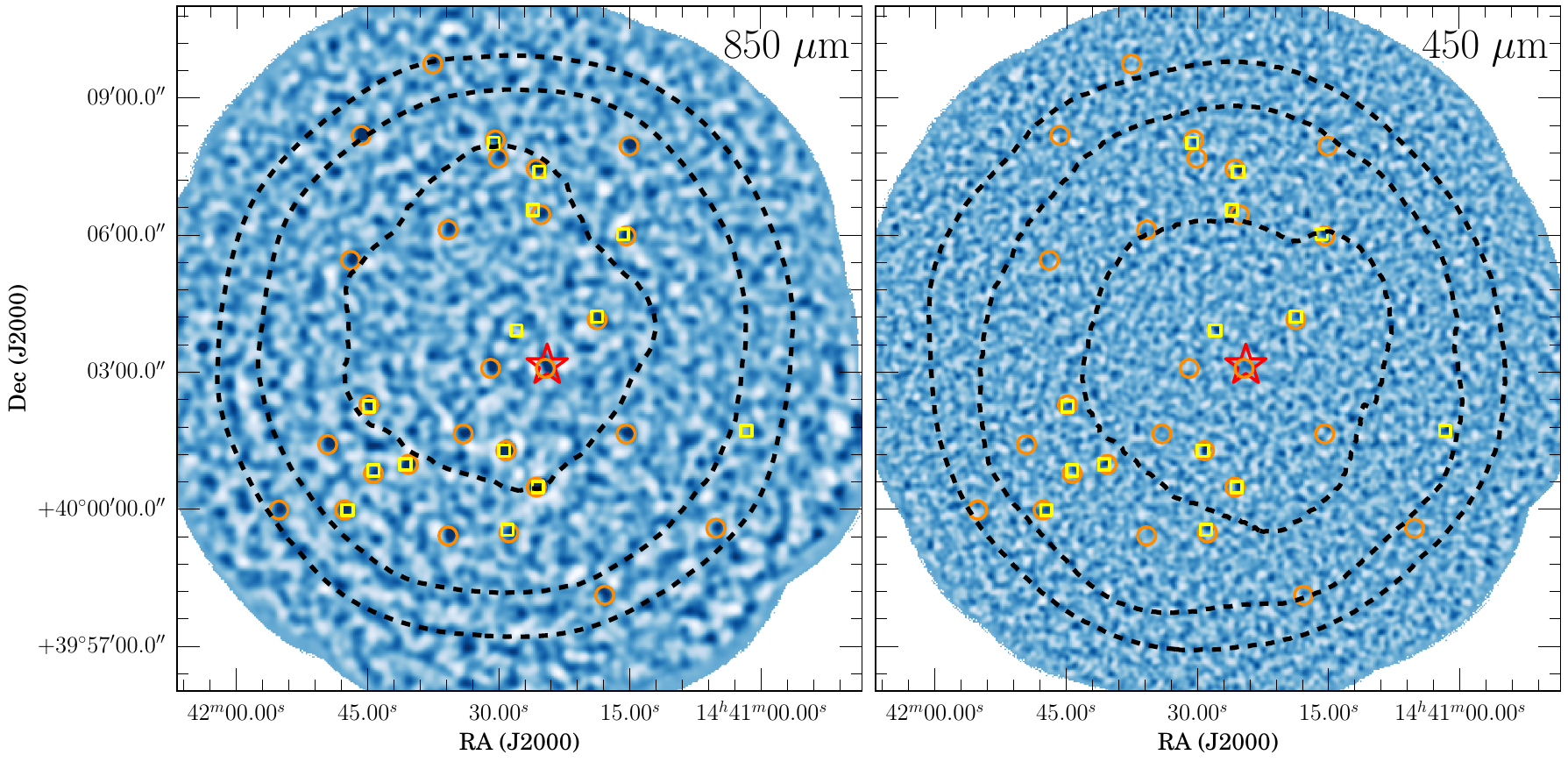}
\caption{SCUBA-2 S/N maps at $850$~$\mu$m (top panel) and $450$~$\mu$m (bottom panel) for the field around the ELAN MAMMOTH-1 (red star).
The maps are shown with a linear scale from -4 to 4. The size of both maps is $15\arcmin \times 15\arcmin$. Orange circles are the 27 
$850$~$\mu$m detections given in Table~\ref{850}, and yellow squares are the 14 $450$~$\mu$m detections given in Table~\ref{450}. 
For both fields, we indicate the noise contours (black dashed) for levels 1.5, 2, and 3$\times$ the central noise ($\sigma_{\rm CN}$; Table~\ref{obs}).
The sizes of the circles and squares correspond to 2$\times$ the beam FWHM of their respective wavelength.}
\label{Maps}
\end{figure*}

The final catalogs were built by keeping every $> 4\sigma$ source in the preliminary catalogs, but also
every $>3\sigma$ source characterized by a $>3\sigma$ counterpart in the other band.
Overall, we discovered 27 sources at 850~$\mu$m and 14 sources at 450~$\mu$m.
In Tables~\ref{850} and \ref{450} we list the information for these sources.
Figure~\ref{Maps} shows the final S/N maps at 850 and 450~$\mu$m for the targeted field with the discovered sources over-plotted. 

\begin{table*}
\begin{center}
\caption{SCUBA-2 850 $\mu$m detected sources around the MAMMOTH-1 Nebula}
\scalebox{1}{
\scriptsize
\setlength\tabcolsep{4pt}
\begin{tabular}{llllrccclccc}
\hline
\hline
 Name                  &  ID$_{850}$           &  R.A.        &  DEC       &    S/N  &  $f_{850}$ & $f_{850}^{\rm Deboosted}$ & $\Delta(\alpha,\delta)$ &  Counterpart	&  S/N$_{\mathrm{c,450}}$  &  $f_{c,450}$ &  $f_{c,450}^{\rm Deboosted}$	\\
                       &               &  (J2000)     &  (J2000)     &         &  (mJy)           &  (mJy)  & ($\arcsec$) &	  ID$_{450}$	    &			    &  (mJy)	 &  (mJy)	      \\
                       (1)&	(2)&	(3)&	(4)&	(5) & (6)&(7)&(8)&(9) & (10) & (11) & (12) \\
\hline
 \multicolumn{9}{l}{ $>4~\sigma$ Sample} \\ 
  SMM J144125.7+400029  & MAM-850.1   &  14:41:25.7  &  +40:00:29  &  16.0  &  21.0$\pm$1.3 & 18.3$\pm$2.8 & 1.0 &  MAM-450.3  &  4.7  &  34$\pm$7  & 14$\pm$7 \\
  SMM J144129.2+400117  & MAM-850.2   &  14:41:29.2  &  +40:01:17  &  15.9  &  18.8$\pm$1.2 & 16.3$\pm$2.7 & 1.1 &  MAM-450.1  &  5.8  &  39$\pm$7  & 21$\pm$8 \\
  SMM J144145.8+400811  & MAM-850.3   &  14:41:45.8  &  +40:08:11  &   7.7  &  13.9$\pm$1.8 &  9.6$\pm$2.1 & 1.6 &  -          &  0.8  &  11$\pm$32 & - \\
  SMM J144140.3+400059  & MAM-850.4   &  14:41:40.3  &  +40:00:59  &   7.4  &  10.0$\pm$1.3 &  6.9$\pm$1.5 & 1.7 &  MAM-450.4  &  4.2  &  36$\pm$8  & 14$\pm$8 \\
  SMM J144144.9+400217  & MAM-850.5   &  14:41:44.9  &  +40:02:17  &   6.9  &   8.9$\pm$1.3 &  6.0$\pm$1.3 & 1.8 &  MAM-450.12 &  3.3  &  28$\pm$8  & 9$\pm$6 \\
  SMM J144125.7+400727  & MAM-850.6   &  14:41:25.7  &  +40:07:27  &   6.2  &   8.1$\pm$1.3 &  5.4$\pm$1.3 & 1.9 &  -  	&  2.4  &  22$\pm$9  & - \\
  SMM J144115.1+400757  & MAM-850.7   &  14:41:15.1  &  +40:07:57  &   5.9  &   9.3$\pm$1.6 &  6.2$\pm$1.6 & 1.9 &  -  	& -0.5  &  -5$\pm$10 & - \\
  SMM J144147.6+395959  & MAM-850.8   &  14:41:47.6  &  +39:59:59  &   5.3  &   8.5$\pm$1.6 &  5.6$\pm$1.7 & 2.1 &  MAM-450.10 &  3.5  &  33$\pm$10 & 11$\pm$7 \\
  SMM J144135.8+395925  & MAM-850.9   &  14:41:35.8  &  +39:59:25  &   5.2  &   7.6$\pm$1.5 &  4.9$\pm$1.6 & 2.1 &  -  	&  0.8  &   7$\pm$9  & - \\
  SMM J144115.4+400559  & MAM-850.10  &  14:41:15.4  &  +40:05:59  &   5.1  &   7.2$\pm$1.4 &  4.6$\pm$1.6 & 2.2 &  MAM-450.8  &  3.6  &  29$\pm$8  & 11$\pm$7 \\
  SMM J144149.6+400125  & MAM-850.11  &  14:41:49.6  &  +40:01:25  &   5.1  &   7.3$\pm$1.5 &  4.6$\pm$1.6 & 2.2 &  -  	&  1.5  &  14$\pm$9  & - \\
  SMM J144130.9+400305  & MAM-850.12  &  14:41:30.9  &  +40:03:05  &   4.9  &   4.5$\pm$0.9 &  2.8$\pm$1.0 & 2.2 &  -          & -0.4  &  -3$\pm$6  & -\\
  SMM J144147.0+400527  & MAM-850.13  &  14:41:47.0  &  +40:05:27  &   4.9  &   6.7$\pm$1.4 &  4.1$\pm$1.5 & 2.2 &  -          &  0.2  &   2$\pm$10 & -\\
  SMM J144124.7+400305  & MAM-850.14  &  14:41:24.7  &  +40:03:05  &   4.9  &   4.6$\pm$0.9 &  2.8$\pm$1.0 & 2.2 &  -          &  2.0  &  11$\pm$6  & -\\
  SMM J144115.4+400139  & MAM-850.15  &  14:41:15.4  &  +40:01:39  &   4.7  &   6.7$\pm$1.4 &  3.9$\pm$1.8 & 2.4 &  -          &  1.0  &   8$\pm$8  & -\\
  SMM J144134.1+400139  & MAM-850.16  &  14:41:34.1  &  +40:01:39  &   4.4  &   5.2$\pm$1.2 &  2.9$\pm$1.5 & 2.5 &  -          &  0.4  &   3$\pm$7  & -\\
  SMM J144155.1+395959  & MAM-850.17  &  14:41:55.1  &  +39:59:59  &   4.3  &   9.8$\pm$2.3 &  5.2$\pm$2.9 & 2.6 &  -          & -0.1  &  -2$\pm$14 & -\\
  SMM J144135.8+400607  & MAM-850.18  &  14:41:35.8  &  +40:06:07  &   4.2  &   5.2$\pm$1.2 &  2.7$\pm$1.5 & 2.7 &  -          &  0.7  &   6$\pm$8  & -\\
  SMM J144130.1+400741  & MAM-850.19  &  14:41:30.1  &  +40:07:41  &   4.1  &   5.4$\pm$1.3 &  2.7$\pm$1.6 & 2.7 &  -          &  0.8  &   7$\pm$9  & -\\
  SMM J144125.2+400627  & MAM-850.20  &  14:41:25.2  &  +40:06:27  &   4.1  &   5.2$\pm$1.3 &  2.7$\pm$1.6 & 2.7 &  -          & -0.5  &  -4$\pm$8  & -\\
  SMM J144128.8+395929  & MAM-850.21  &  14:41:28.8  &  +39:59:29  &   4.0  &   5.7$\pm$1.4 &  2.8$\pm$1.7 & 2.8 &  MAM-450.9  &  3.5  &  29$\pm$8  & 10$\pm$6 \\
  SMM J144117.9+395807  & MAM-850.22  &  14:41:17.9  &  +39:58:07  &   4.0  &	8.1$\pm$2.0 &  4.0$\pm$2.4 & 2.8 &  -	       &  1.5  &  16$\pm$11 & -\\
  SMM J144105.2+395935  & MAM-850.23  &  14:41:05.2  &  +39:59:35  &   4.0  &	8.9$\pm$2.2 &  4.4$\pm$2.7 & 2.8 &  -	       &  1.2  &  15$\pm$12 & -\\
  SMM J144137.6+400945  & MAM-850.24  &  14:41:37.6  &  +40:09:45  &   4.0  &  10.5$\pm$2.6 &  5.2$\pm$3.2 & 2.8 &  -	       & -0.1  &  -2$\pm$20 & -\\
 \multicolumn{9}{l}{ $>3~\sigma$ Sample with $>3~\sigma$ Counterparts at 450\,$\mu$m} \\ 	     
  SMM J144144.3+400047  & MAM-850.25  &  14:41:44.3  &  +40:00:47  &   3.9  &   5.7$\pm$1.5 &  2.8$\pm$1.7 & 2.8 & MAM-450.14  & 3.0   & 26$\pm$9  & 9$\pm$6 \\
  SMM J144130.4+400805  & MAM-850.26  &  14:41:30.4  &  +40:08:05  &   3.9  &   4.7$\pm$1.2 &  2.3$\pm$1.4 & 2.8 & MAM-450.11  & 3.4   & 32$\pm$9  & 11$\pm$7 \\
  SMM J144118.7+400409  & MAM-850.27  &  14:41:18.7  &  +40:04:09  &   3.7  &   4.1$\pm$1.1 &  2.0$\pm$1.2 & 2.9 & MAM-450.13  & 3.1   & 21$\pm$7  & 7$\pm$5 \\
\hline
\hline
\end{tabular}
}
\flushleft{(1) Name of source; (2) our identification for the source; (3) and (4) right ascension and declination in J2000 coordinates; (5) S/N at 850~$\mu$m; 
(6) flux at 850~$\mu$m; (7) Deboosted flux obtained using the mean curve shown in Figure~\ref{Flux_boosting}; (8) positional error as derived in section~\ref{pos_err}; 
(9) our identification for the counterpart; 
(10) S/N of the counterpart or the S/N measured at the peak position at 450~$\mu$m; (11) flux for
the counterpart or flux measured at the peak position at 450~$\mu$m; (12) Deboosted flux for the counterpart obtained using the mean curve 
shown in Figure~\ref{Flux_boosting}.}
\label{850}
\end{center}
\end{table*}

\begin{table*}
\begin{center}
\caption{SCUBA-2 450 $\mu$m detected sources around the MAMMOTH-1 Nebula}
\scalebox{1}{
\scriptsize
\setlength\tabcolsep{4pt}
\begin{tabular}{llllrccclccc}
\hline
\hline
 Name                  &  ID$_{450}$           &  R.A.        &  DEC       &    S/N  &  $f_{450}$      & $f_{450}^{\rm Deboosted}$ & $\Delta(\alpha,\delta)$  &   Counterpart  &  S/N$_{c,850}$  &  $f_{c,850}$ &  $f_{c,850}^{\rm Deboosted}$         \\
                       &               &  (J2000)     &  (J2000)     &         &  (mJy)                & (mJy) & ($\arcsec$)  &	ID$_{850}$	   &		&   (mJy) &   (mJy)		  \\
                       (1)&	(2)&	(3)&	(4)&	(5) & (6)&(7)&(8)&(9) & (10) & (11) & (12)\\
\hline
 \multicolumn{12}{l}{ $>4~\sigma$ Sample} \\ 
 SMM J144129.4+400117  &  MAM-450.1  &  14:41:29.4  &  +40:01:17  &  5.8  &  38$\pm$7  & 21$\pm$8 & 1.4 &  MAM-850.2  &  15.9  &  18.8$\pm$1.2 & 16.3$\pm$2.7 \\
 SMM J144125.4+400723  &  MAM-450.2  &  14:41:25.4  &  +40:07:23  &  4.8  &  42$\pm$9  & 18$\pm$9 & 1.8 &  -          &   5.0  &   6.6$\pm$1.3 & - \\
 SMM J144125.5+400029  &  MAM-450.3  &  14:41:25.5  &  +40:00:29  &  4.7  &  34$\pm$7  & 14$\pm$7 & 1.9 &  MAM-850.1  &  16.0  &  21.0$\pm$1.3 & 18.3$\pm$2.8 \\
 SMM J144140.7+400059  &  MAM-450.4  &  14:41:40.7  &  +40:00:59  &  4.2  &  35$\pm$8  & 14$\pm$8 & 2.0 &  MAM-850.4  &	  7.4  &  10.0$\pm$1.3 &  6.9$\pm$1.5 \\
 SMM J144126.1+400633  &  MAM-450.5  &  14:41:26.1  &  +40:06:33  &  4.2  &  35$\pm$8  & 13$\pm$8 & 2.0 &  -          &  -0.3  &  -0.4$\pm$1.3 & - \\
 SMM J144128.0+400355  &  MAM-450.6  &  14:41:28.0  &  +40:03:55  &  4.1  &  23$\pm$6  &  9$\pm$5 & 2.0 &  -          &  -0.6  &  -0.5$\pm$0.9 & - \\
 SMM J144101.7+400143  &  MAM-450.7  &  14:41:01.7  &  +40:01:43  &  4.0  &  41$\pm$10 & 16$\pm$9 & 2.1 &  -          &   0.3  &   0.6$\pm$1.9 & - \\
 \multicolumn{12}{l}{ $>3~\sigma$ Sample with $>3~\sigma$ Counterparts at 850\,$\mu$m} \\ 	    								      
 SMM J144115.8+400601  &  MAM-450.8   &  14:41:15.8  &  +40:06:01  &  3.6  &  29$\pm$8  & 11$\pm$7 & 2.2 &  MAM-850.10  &  5.1  &  7.2$\pm$1.4  & 4.6$\pm$1.6 \\
 SMM J144129.0+395933  &  MAM-450.9   &  14:41:29.0  &  +39:59:33  &  3.5  &  29$\pm$8  & 10$\pm$6 & 2.2 &  MAM-850.21  &  4.0  &  5.7$\pm$1.4  & 2.8$\pm$1.7 \\
 SMM J144147.3+395959  &  MAM-450.10  &  14:41:47.3  &  +39:59:59  &  3.5  &  33$\pm$10 & 11$\pm$7 & 2.3 &  MAM-850.8   &  5.4  &  8.5$\pm$1.6  & 5.6$\pm$1.7 \\
 SMM J144130.6+400801  &  MAM-450.11  &  14:41:30.6  &  +40:08:01  &  3.4  &  32$\pm$9  & 11$\pm$7 & 2.3 &  MAM-850.26  &  3.9  &  4.7$\pm$1.2  & 2.3$\pm$1.4 \\
 SMM J144144.9+400215  &  MAM-450.12  &  14:41:44.9  &  +40:02:15  &  3.3  &  28$\pm$8  &  9$\pm$6 & 2.3 &  MAM-850.5   &  6.9  &  8.9$\pm$1.3  & 6.0$\pm$1.3 \\
 SMM J144118.7+400413  &  MAM-450.13  &  14:41:18.7  &  +40:04:13  &  3.1  &  21$\pm$7  &  7$\pm$5 & 2.4 &  MAM-850.27  &  3.7  &  4.1$\pm$1.1  & 2.0$\pm$1.2 \\
 SMM J144144.3+400051  &  MAM-450.14  &  14:41:44.3  &  +40:00:51  &  3.0  &  26$\pm$9  &  9$\pm$6 & 2.4 &  MAM-850.25  &  3.9  &  5.7$\pm$1.5  & 2.8$\pm$1.7 \\
\hline
\hline
\end{tabular}
}
\flushleft{(1) Name of source; (2) our identification for the source; (3) and (4) right ascension and declination in J2000 coordinates; (5) S/N at 450~$\mu$m; 
(6) flux at 450~$\mu$m; (7) Deboosted flux obtained using the mean curve shown in Figure~\ref{Flux_boosting}; 
(8) positional error as derived in section~\ref{pos_err}; (9) our identification for the counterpart; 
(10) S/N of the counterpart or the S/N measured at the peak position at 850~$\mu$m; (11) flux for
the counterpart or flux measured at the peak position at 850~$\mu$m; (12) Deboosted flux for the counterpart obtained using the mean curve 
shown in Figure~\ref{Flux_boosting}.}
\label{450}
\end{center}
\end{table*}

\subsection{Reliability of Source Extraction}
\label{jackknife}

To determine the number of spurious sources that could affect our catalogues, we proceeded as follows.
First, we applied the source extraction algorithm to the inverted maps. We found 2 and 1 detections at $>4\sigma$
at 850 and 450 $\mu$m, respectively.
Second, we constructed true noise maps, we applied the source extraction algorithm, and checked the number of 
detections with $>4\sigma$. To obtain true noise maps we used the jackknife resampling technique. Specifically, we subtracted
two maps obtained by co-adding roughly half of the data for each band. In this way, any real source in the maps is subtracted
irrespective of its significance. The residual maps are thus source-free noise maps.
To account for the difference in exposure time, we scaled these true noise maps by a factor of 
$\sqrt{t1\times t2}/(t1+t2)$, with t1 and t2 being the exposure time of each pixel from the two maps.
These jackknife maps are characterized by a central noise of 0.88 and 5.39 mJy/beam, respectively at 850~$\mu$m and 450~$\mu$m, in agreement 
with the noise in the science data. 
By applying the source extraction algorithm to these maps, we found 1 and 4 detections at $>4\sigma$
at $850$ and $450$~$\mu$m, respectively.
We thus expect a similar number of spurious sources in our $>4\sigma$ source catalogs. 

Further, we tested the number of spurious detections for the $3\sigma$ sources identified as having a counterpart in the 
other bandpass (lower portion of Table~\ref{850} and \ref{450}) by using once again the jackknife maps. Specifically,
from this maps we extracted sources between $3$ and $4\sigma$, and cross-correlated them with the detections in the real data
at the other wavelength. We found that none of such spurious sources matched a detection in the real maps.
We then repeated the test $1000$ times by randomizing the position of the spurious sources within the effective area of our observations.
On average we found $0.3$ and $0.7$ spurious sources at $850$ and $450$~$\mu$m, indicating that sources selected at the $>3\sigma$ level
in both bandpasses are even more reliable than $>4\sigma$ sources selected in only one bandpass.

Overall these tests suggest that -- most likely -- the sources at $450$~$\mu$m without a detection at $850$~$\mu$m are spurious for our observations. 
For the sake of completeness, we decided 
to list all the sources in our catalogs. As it will be clear from our analysis, our conclusions are not affected.

\subsection{Completeness Tests}

\begin{figure}
\centering
\includegraphics[width=0.98\columnwidth]{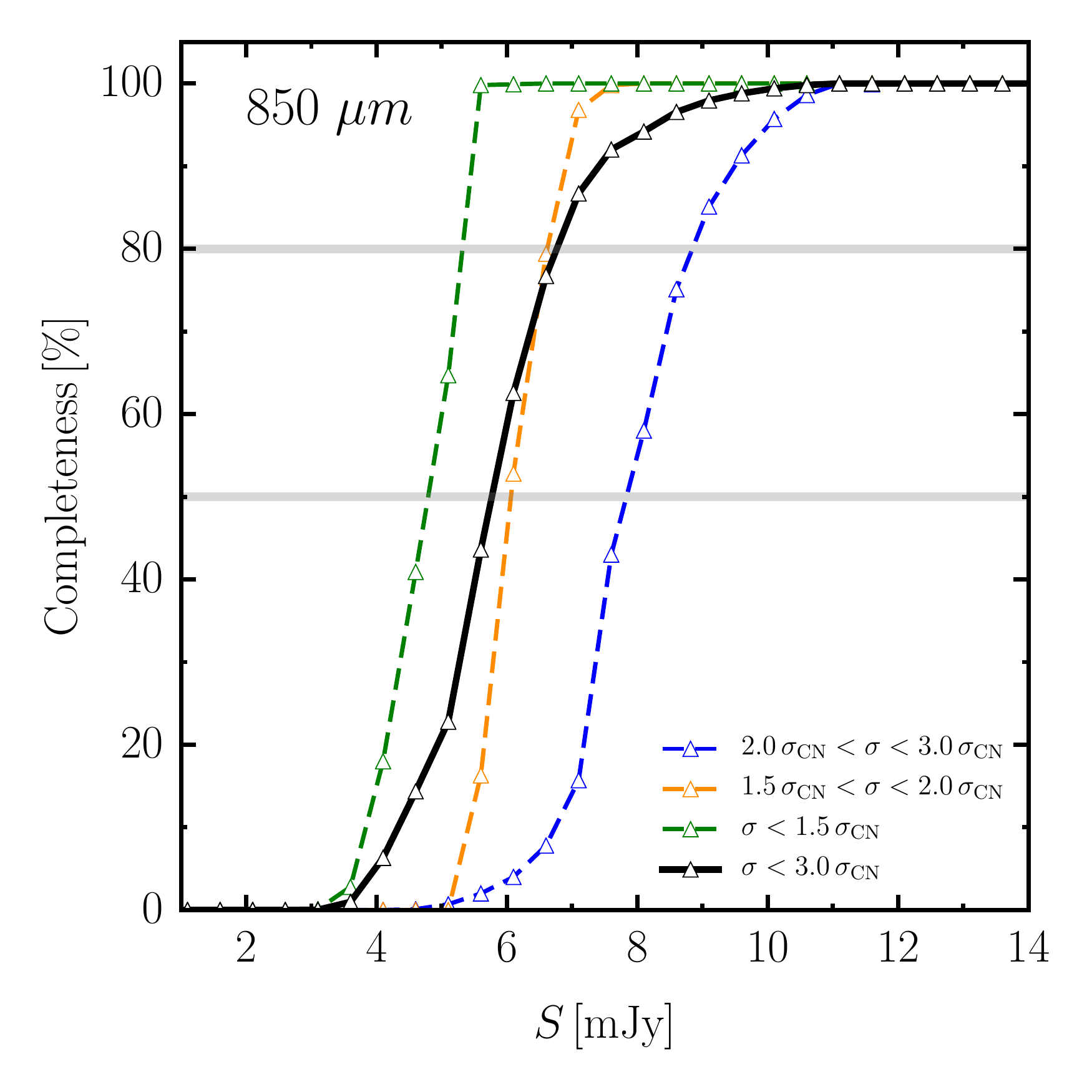}\\
\includegraphics[width=0.98\columnwidth]{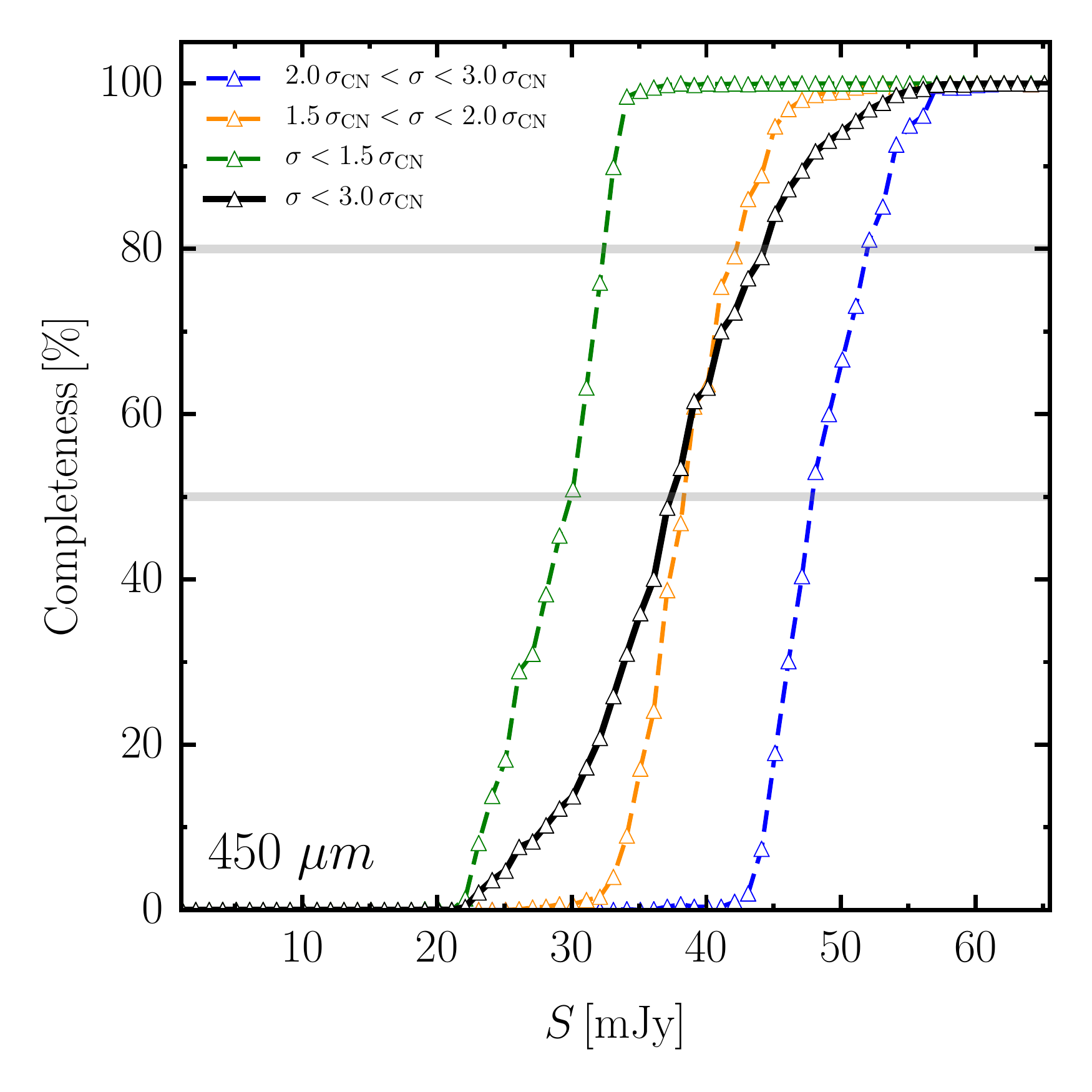}
\caption{Top: completeness at $850$~$\mu$m versus flux for different portions of the map, i.e. 
$2\sigma_{\rm CN}<\sigma<3\sigma_{\rm CN}$, 
$1.5\sigma_{\rm CN}<\sigma<2\sigma_{\rm CN}$, $\sigma<1.5\sigma_{\rm CN}$, and $<3\sigma_{\rm CN}$ (see Fig.~\ref{Maps}).
Bottom: same as above, but fot the $450$~$\mu$m dataset.}
\label{completeness}
\end{figure}

We tested at which flux our data can be considered complete.
We proceeded as follows. We took the true noise maps introduced in section~\ref{jackknife}, 
and populated them with mock sources of a given flux and placed at random positions.
We have then extracted the sources, considering them as recovered if the detection is above 4$\sigma$ and
within the beam area. Specifically, we injected sources with flux from 0.1 to 25.1 mJy (0.1 to 80.1 mJy) with a step of
0.5 mJy (1.0 mJy) for 850 (450) $\mu$m. For each step in flux we iterate the extraction by introducing 1000 sources.
To fully characterize the completeness in the whole extent of the maps, we repeated this procedure for areas 
of the images characterized by different depths, i.e. $<3\sigma_{\rm CN}$, $2\sigma_{\rm CN}<\sigma<3\sigma_{\rm CN}$, 
$1.5\sigma_{\rm CN}<\sigma<2\sigma_{\rm CN}$, and $\sigma<1.5\sigma_{\rm CN}$ (see Fig.~\ref{Maps}).
Figure~\ref{completeness} shows the results of the tests.
For the whole area with $<3\sigma_{\rm CN}$, the $50\%$ completeness is at 5.8 and 37 mJy at 850 and 450~$\mu$m, respectively, and
the $80\%$ is at 6.8 and 44 mJy, respectively.
As expected the central portion of the maps with $\sigma<1.5\sigma_{\rm CN}$  have a better sensitivity, with 
the $50\%$ completeness being around 4.8 and 30 mJy at 850 and 450~$\mu$m, respectively, and
the $80\%$ being around 5.3 and 32 mJy, respectively.

\section{Number Counts}
\label{NC}

\begin{figure}
\centering
\includegraphics[width=0.95\columnwidth]{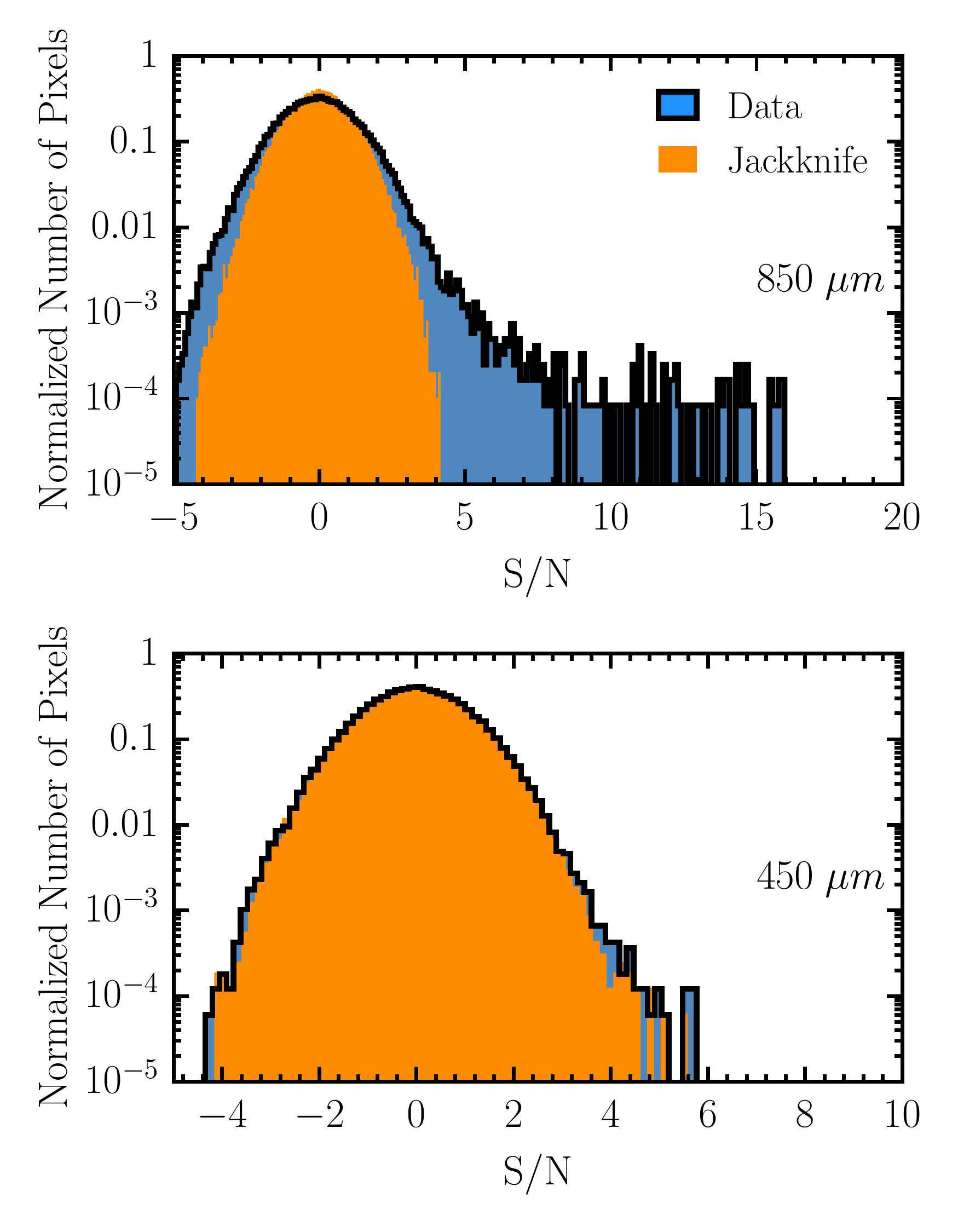}
\caption{Normalized histograms of the S/N values for the pixels within the portions of the 850 and 450~$\mu$m maps characterized by less than three times the central noise. 
The orange and blue histograms indicate the distributions of the jackknife maps, and of the data, respectively.
The jackknife maps represents the pixel noise distributions which dominates at low S/N (see Section~\ref{jackknife} for details). The data (especially at 850~$\mu$m) 
shows excesses at high S/N where sources contribute to the distributions. The matched filter technique introduces residual troughs around bright detections, which are visible
here as negative excesses.}
\label{histos}
\end{figure}

In this section we determine the pure source number counts at 850 and 450~$\mu$m around the ELAN MAMMOTH-1, and 
estimate the underlying counts models for each wavelength.
A precise measurement of the galaxy number counts needs an accurate 
estimate of the number of spurious sources contaminating the counts. For this purpose, we followed the 
procedure in \citet{TC2013a,TC2013b}, and use the jackknife maps produced in Section~\ref{jackknife} to assess how many
spourious sources affect the counts. 
As a first step, in Figure~\ref{histos} we show the S/N histograms of the true noise maps (orange shading) and the 
signal maps (blue shading with black edge). The excess signals with respect to the pure noise distribution are from
real astronomical sources. On the other hand, the negative excesses are due to the negative throughs of the matched-filter PSF.
From these histograms it is well evident that the 450~$\mu$m data are less sensitive and more affected by the presence of
spurious sources (as already noted in Section~\ref{jackknife}).

In contrast to what done with the catalogs in Section~\ref{catalogs} -- where we have selected only detections with S/N$>4$ or with 
S/N$>3$ at both wavelengths -- we lowered our detection threshold to $2\sigma$.
Indeed, as the positional information is not relevant for number counts analyses, the detection threshold
can be lowered to explore statistically significant positive excesses (e.g., \citealt{TC2013b}).
We thus use the preliminary catalogues produced in Section~\ref{catalogs}, and additionally ran the source extraction
algorithm on the true noise maps down to S/N$=2$.

\begin{figure}
\centering
\includegraphics[width=0.98\columnwidth]{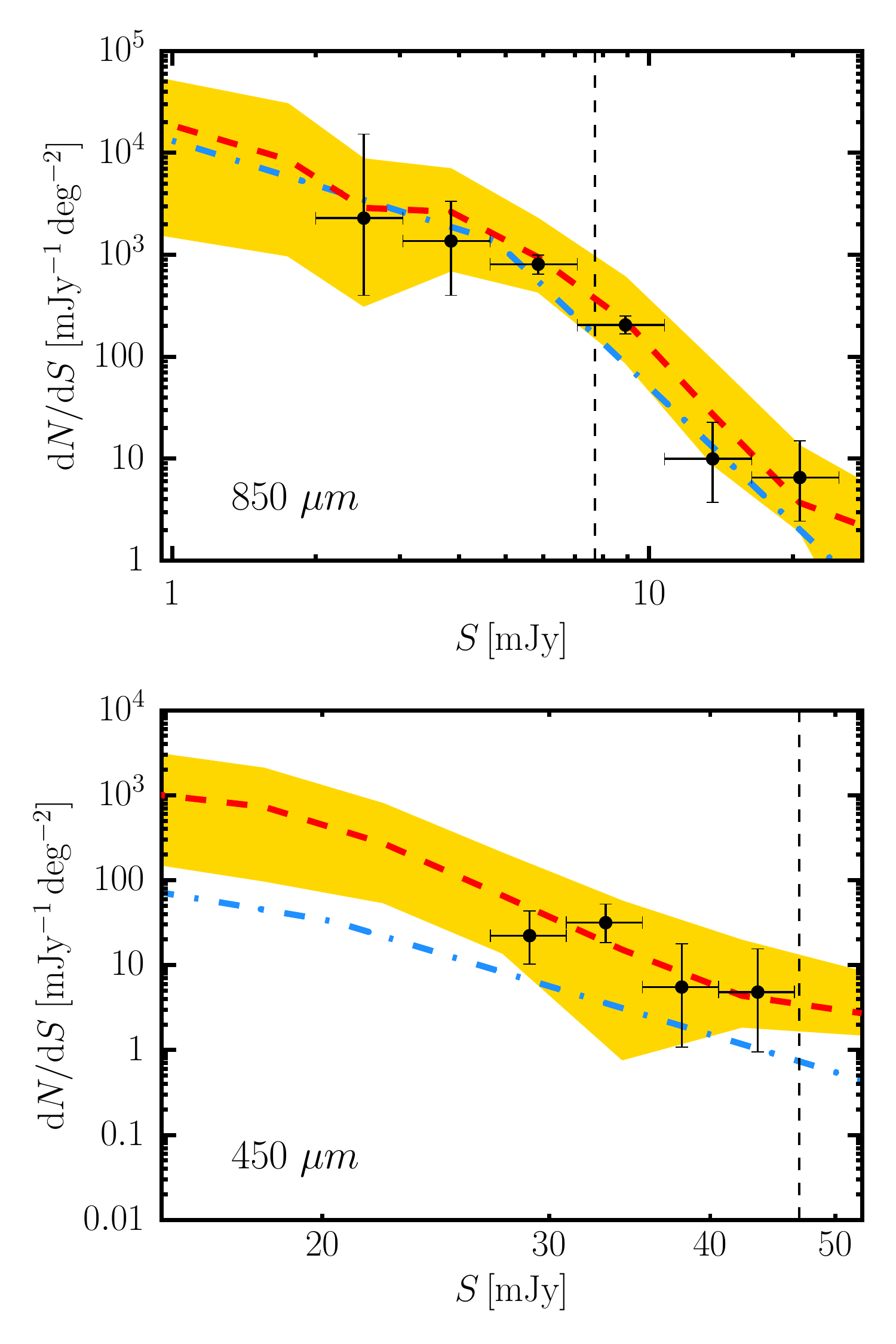}
\caption{Pure source differential number counts (black data-point) at 850 and 450~$\mu$m around the ELAN MAMMOTH-1 compared to the simulated mean counts 
(red dashed line). The yellow shadings represent the 90\% confidence range obtained from 500 realizations of the blue dot-dashed curves. The blue dot-dashed curves are the
final adopted underlying models for the Monte Carlo simulations (see section~\ref{NC}), and represent the true number counts. The dashed vertical lines
indicate the mean $4\sigma$ within the effective area. The horizontal errorbars for the data-points indicate the width of each flux bin.}
\label{counts_MCMC}
\end{figure}

The pure source differential number counts are then obtained as follows.
First, for each extracted source in the signal map we calculated the number density by inverting the detectable area, which is the portion of the
field-of-view with noise level low enough to allow the detection of the source. Second, we obtained the 
raw number counts by summing up the number densities of the sources within each flux bin. Finally,
to get to the pure source differential number counts, we subtracted the number counts similarly obtained from the 
true noise maps, if any, from the counts obtained from the signal maps.
Figure~\ref{counts_MCMC} shows the obtained pure source differential number counts (black data points) for both 850 (top panel) and 
450~$\mu$m (lower panel).

To obtain the underlying counts models we ran Monte Carlo simulations following e.g., \citet{TC2013a,TC2013b}.
First, we create a simulated image by randomly injecting mock sources onto the jackknife maps. The mock sources are 
drawn from an assumed model and convolved with the PSFs. 
For the counts models we use a broken power-law of the form

\begin{equation}
\label{eqn:diff_counts}
  \frac{dN}{dS} = \left\{
  \begin{array}{l l}
    {N_0}\left(\frac{S}{S_0}\right)^{-\alpha}  & \quad \text{if $S \leq S_0$}\\
    {N_0}\left(\frac{S}{S_0}\right)^{-\beta}  & \quad \text{if $S > S_0$}\\
  \end{array} \right.,
\end{equation} 
and started from a fit to the observed counts.
As faintest fluxes for our models, we use the fluxes at which the integrated flux density agrees with the 
values for the extragalactic background light (EBL; e.g., \citealt{Puget1996}).

After obtaining a mock map, we ran the source extraction algorithm and computed the number
counts in exactly the same way as done with the real data. 
We then calculated the ratio between the recovered counts and the input model, which reflects the Eddington
bias (\citealt{Eddington1913}), and then applied this ratio to the observed counts to correct
for this bias. A $\chi^2$ fit is performed to the corrected observed counts using the broken power-law
to get the normalization and power-law indices, which are then used in the next iteration. 
This iterative process was terminated once the input model agreed with the corrected counts at the $1\sigma$ level.
Given the low number of data-points at 450~$\mu$m, we only fitted the normalization and the bright-end slope at this wavelength.
We fixed $S_0$ and $\alpha$ to the values in \citet{TC2013b}.

To test the reliability of our results, we have then created 500 realizations of simulated maps using as input the model curves determined 
through the Monte Carlo simulations, and calculated the pure source number counts for each of them.
In Figure~\ref{counts_MCMC} we show the results of the Monte Carlo simulations and compare them to the data.
We give the derived underlying counts models (blue dot-dashed lines), the mean counts (red dashed lines), and the 90\%
confidence range of the 500 realization (yellow).
The 500 realizations well match the pure source number counts within the uncertainties. We can then apply the ratio between 
the mean number counts and the input model to correct our data, and thus obtain the true differential number counts (see Section~\ref{true_number_counts}).
Table~\ref{MC_input} summarizes the parameters of the obtained underlying count models at both 850 and 450~$\mu$m.

\begin{table}
\begin{center}
\caption{The true number counts curves at 450 and 850~$\mu$m from the Monte Carlo simulations
using the broken power-law shown in eqn.~\ref{eqn:diff_counts}.}
\begin{tabular}{cccccc}
\hline
\hline
 Wavelengths  &    N$_0$  &  S$_0$  &  $\alpha$  &  $\beta$  \Tstrut\\
($\mu$m)      &    (mJy$^{-1}$~deg$^{-2}$) & (mJy) &  &      \Bstrut\\
\hline
         450  &        33  &   20.4 &	   2.53  &     4.56  \Tstrut\\
         850  &      1380  &   4.76  &	   1.45  &     4.44  \Bstrut\\
\hline
\hline
\end{tabular}
\label{MC_input}
\end{center}
\end{table}

%% OLD VALUES
%         450  &       284  &   23.39 &	   1.99  &     4.19  \\
%         850  &      1380  &   4.76  &	   1.45  &     4.44  \\

\section{Flux boosting estimates}
\label{fluxBoost}

\begin{figure}
\centering
\includegraphics[width=0.98\columnwidth]{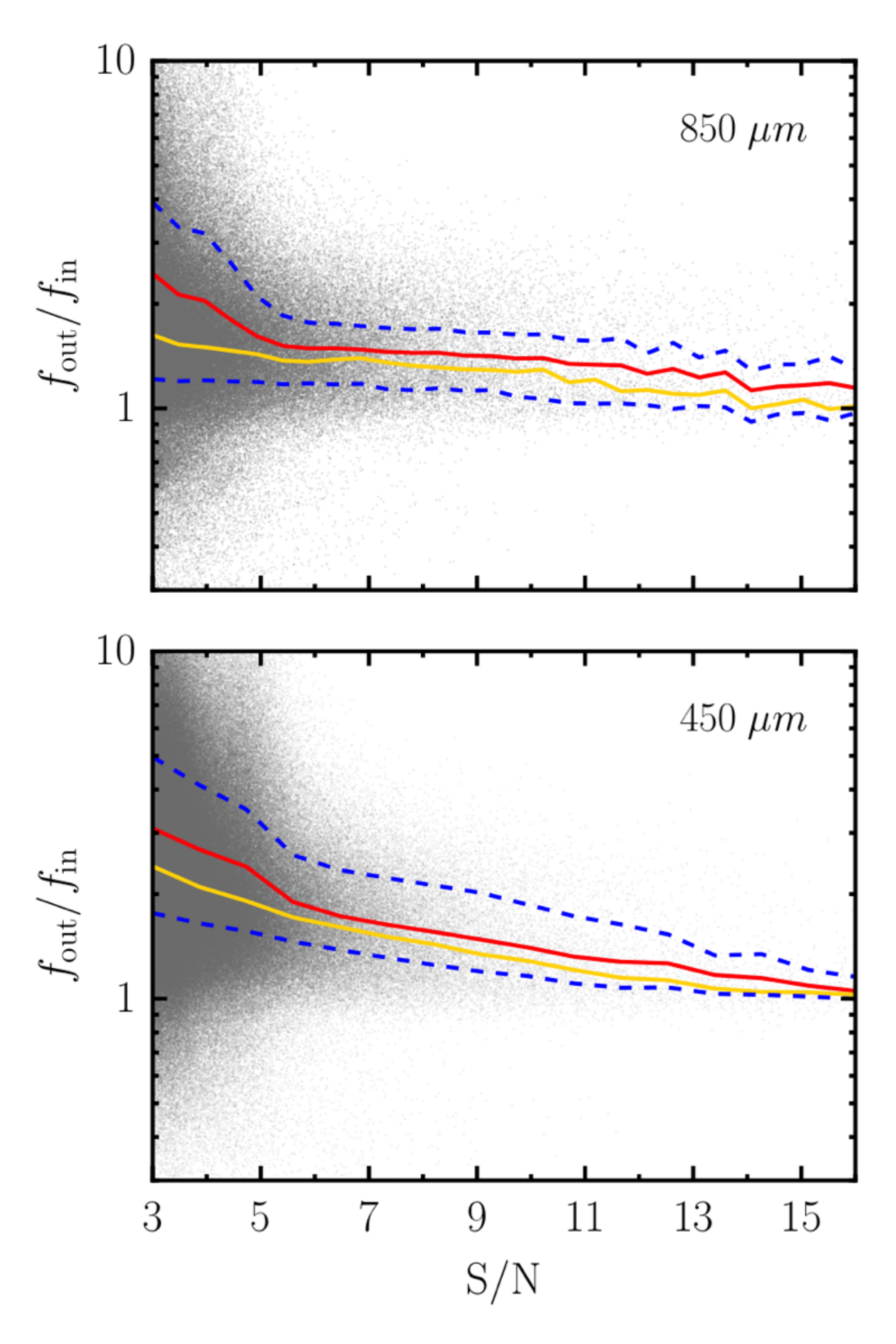}
\caption{Ratio between the fluxes of the detected sources and the injected sources from the 500 realizations of the estimated underlying counts models
(section~\ref{NC}) as a function of the S/N of the detections. The gray dots are $\sim100,000$ simulated data-points. We show the mean (red) 
and median (yellow) values
of the flux ratio in different S/N bins. The blue dashed curves enclose the $1\sigma$ range relative to the mean curve. The test is shown for both 850 (top)
and 450~$\mu$m (bottom).}
\label{Flux_boosting}
\end{figure}

With the Monte Carlo simulations we found a systematic flux/count boost, which we 
characterized by comparing the flux of the injected mock sources with the 
detections. In particular, we selected the brightest input source located within the beam area of each of the 
$>3\sigma$ recovered sources, and computed the flux ratio.
In Figure~\ref{Flux_boosting} we show this test as a function of S/N for both wavelengths. 
We plot the mean (red) and the median (yellow) values of the flux boosting,
together with the $1\sigma$ ranges (blue) relative to the mean values.
At S/N$=4$, the estimated median flux boosting is 2.0 and 1.5 at 450 and 850~$\mu$m, respectively.
This values are in agreement within the uncertainties with similar previous studies conducted with SCUBA-2 (e.g., \citealt{Casey2013,TC2013a}).
We then corrected the observed fluxes for the catalogs obtained in Section~\ref{catalogs} using the median curves, 
and listed the de-boosted fluxes in Tables~\ref{450} and \ref{850}.
This flux boost is usually found in previous SCUBA studies (e.g., \citealt{Wang2017}) and 
it is ascribed to the so-called Eddington bias (\citealt{Eddington1913}).

\section{Positional uncertainties}
\label{pos_err}

\begin{figure}
\centering
\includegraphics[width=0.98\columnwidth]{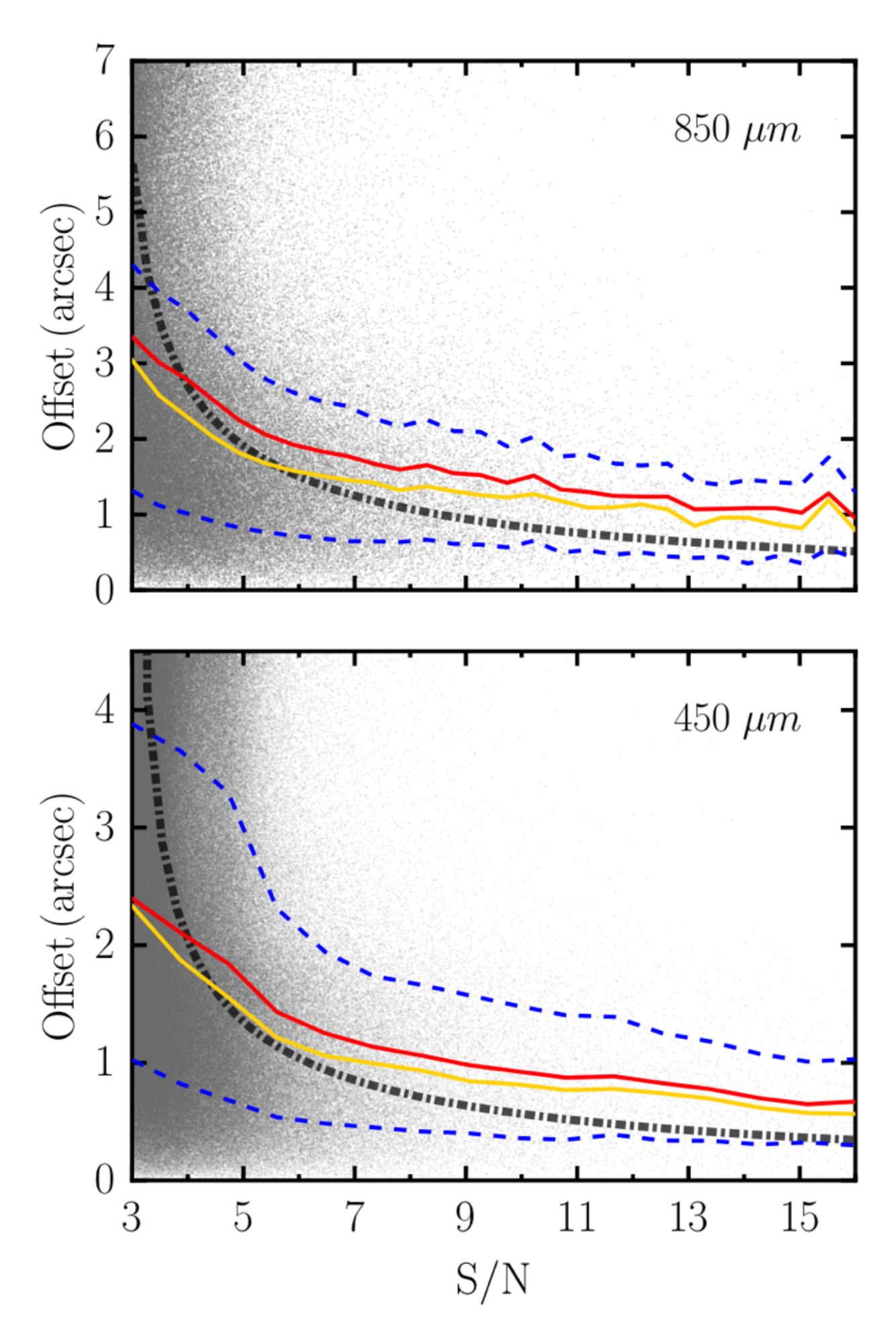}
\caption{Positional offset between the detected sources and the injected sources from the 500 realizations of the estimated underlying counts models
(section~\ref{NC}) as a function of the S/N of the detections. The gray dots are $\sim100,000$ simulated data-points. We show the mean (red) 
and median (yellow) values
of the positional offsets in different S/N bins. The blue dashed curves enclose the $1\sigma$ range relative to the mean curve. 
The test is shown for both 850 (top) and 450~$\mu$m (bottom). The dot-dashed black lines indicate the predictions from \citet{Ivison2007} based on the LESS
sample.}
\label{pos_off}
\end{figure}

Using the same Monte Carlo simulations and the same algorithm to find counterparts in the injected and
recovered catalogs, we can estimate the positional offset between the location of the
injected and the recovered sources. 
Figure~\ref{pos_off} shows this test at both 450 and 850~$\mu$m. At S/N$\lesssim5$, 
there is a large scatter, suggesting positional uncertainties of the order of $\gtrsim1.7$ and $\gtrsim2.2$~arcsec, respectively
for 450 and 850~$\mu$m. 
At larger S/N the uncertainty is lower, down to 1 arcsec for sources as strong as the brightest objects in our 850~$\mu$m catalog (S/N$\approx16$).
At 450~$\mu$m -- characterized by a $\approx 1.5\times$ smaller beam -- the positional uncertainties are
slightly smaller.
These results well agrees -- within the uncertainties -- with the predicted positional offset based on the LESS sample (dashed black line;
equation B22 in \citealt{Ivison2007}). 
Based on this test, we can then 
assign the mean value of the offsets as positional uncertainty to the 
detections listed in Tables~\ref{450} and \ref{850}.

%--------------------------------------------------------------------
\section{Results}
\label{sec:results}
%--------------------------------------------------------------------

\subsection{True Number Counts}
\label{true_number_counts}

\begin{figure*}
\centering
\includegraphics[width=0.95\textwidth]{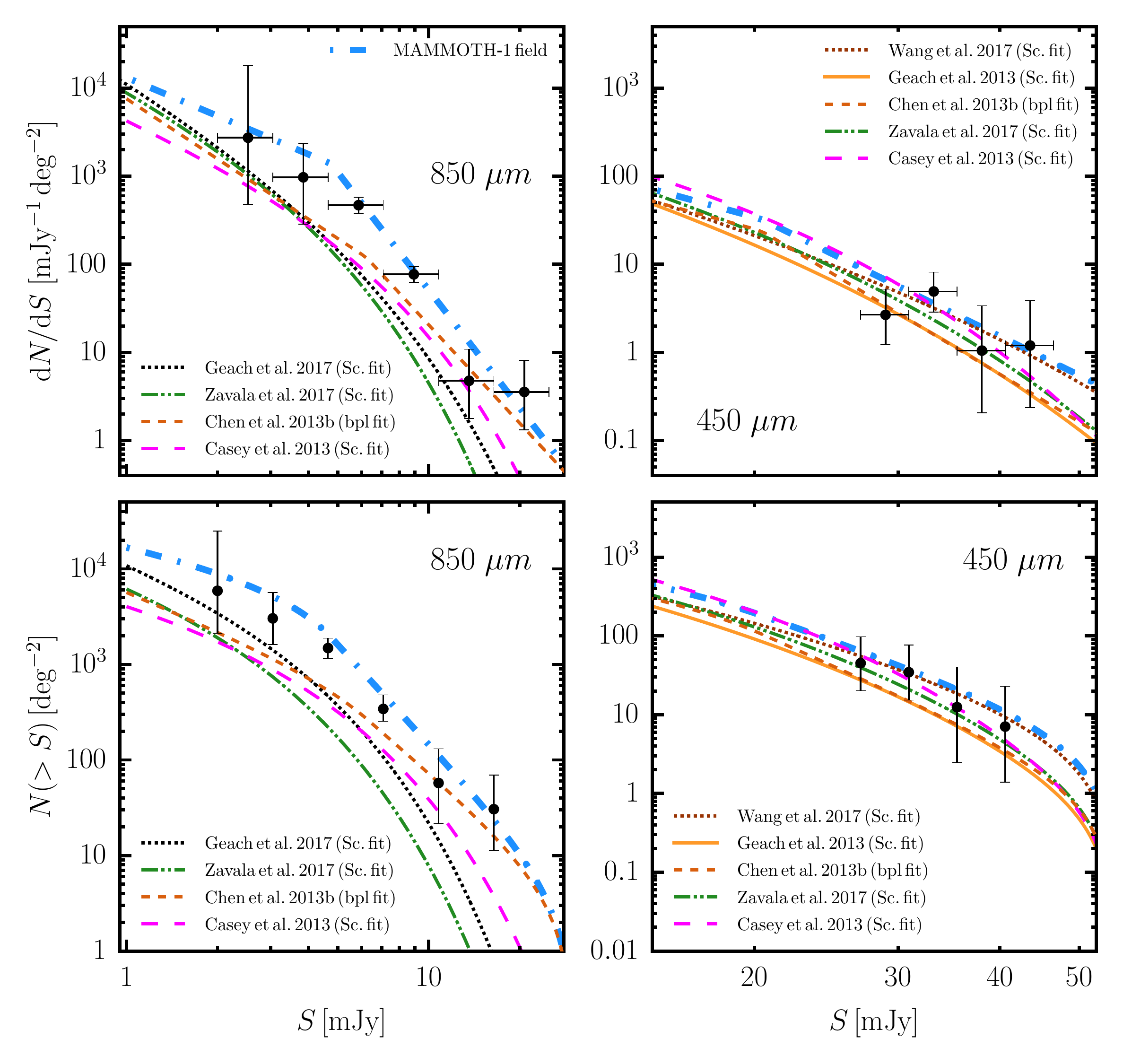}
\caption{Comparison of the SCUBA-2 MAMMOTH-1 differential number counts at 850~$\mu$m (top left) and 450~$\mu$m (top right), and the cumulative number counts at 
850~$\mu$m (bottom left) and 450~$\mu$m (bottom right) with respect to estimates for blank fields in the literature
(\citealt{TC2013b,Casey2013,Geach2013,Wang2017,Geach2017,Zavala2017}). The black filled symbols are the corrected counts following our Monte Carlo simulations 
(see section~\ref{NC}),
and the blue dot-dashed curves are our true number counts curves from Table~\ref{MC_input}. 
When comparing to blank fields, our data for the MAMMOTH-1 field thus show higher counts at 850~$\mu$m, 
but not at 450~$\mu$m.}
\label{true_counts}
\end{figure*}

Figure~\ref{true_counts} presents the corrected differential and cumulative number counts at both 450 and 850~$\mu$m 
for the effective area
of our observations, together with the derived underlying broken power-law (bpl) models (blue dot-dashed lines).
As explained in Section~\ref{NC}, these true number counts have been obtained by dividing the pure source counts by 
the ratio between the mean number counts and the input models.
We list the values of our corrected data-points in Table~\ref{table:counts}.

\begin{table}
\begin{center}
\caption{The true number counts at 450 and 850~$\mu$m.}
\begin{tabular}{cccc}
\hline
\hline
\multicolumn{4}{c}{850~$\mu$m}\Tstrut\Bstrut\\
\hline
$S_{850}$  &    $dN/dS^{a}$  &  $S_{850}$  &  $N(>S)^{b}$   \Tstrut\\
(mJy)      &    (mJy$^{-1}$~deg$^{-2}$) & (mJy) &  \Bstrut\\
\hline
 2.5 	   &	2700$^{+15500}_{-2300}$   &  2.0      &  5900$^{+18900}_{-3800}$	   \Tstrut	\\
 3.8 	   &	1000$^{+1400}_{-700}$     &  3.1      &  3000$^{+2600}_{-1400}$	   \Tstrut	\\
 5.9 	   &	470$^{+107}_{-95}$        &  4.6      &  1500$^{+400}_{-300}$	   \Tstrut	\\
 8.9 	   &	77$^{+17}_{-14}$          &  7.1      &  340$^{+140}_{-90}$	   \Tstrut	\\
 13.6	   &	5$^{+6}_{-3}$             &  10.8     &  57$^{+73}_{-36}$	   \Tstrut	\\
 20.7	   &	4$^{+5}_{-2}$             &  16.4     &  30$^{+40}_{-20}$	   \Tstrut\Bstrut\Bstrut	\\
\hline
\hline
\multicolumn{4}{c}{450~$\mu$m}\Tstrut\Bstrut\\
\hline
$S_{450}$  &    $dN/dS^{a}$  &  $S_{450}$  &  $N(>S)^{b}$   \Tstrut\\
(mJy)      &    (mJy$^{-1}$~deg$^{-2}$) & (mJy) &  \Bstrut\\
\hline
 29.0      &    3$^{+3}_{-1}$        &  27.0  & 45$^{+50}_{-25}$  \Tstrut      \\  	       
 33.2	   &	5$^{+3}_{-2}$        &  30.9  & 35$^{+40}_{-20}$  \Tstrut      \\
 38.0	   &	1.1$^{+2.3}_{-0.8}$  &  35.4  & 12$^{+28}_{-10}$ \Tstrut	   \\
 43.6	   &	1.2$^{+2.7}_{-1.0}$  &  40.6  &  7$^{+16}_{-6} $ \Tstrut\Bstrut\Bstrut	    \\
\hline
\hline
\multicolumn{4}{l}{$^{a}$ {\footnotesize Differential number counts.}}\Tstrut\\
\multicolumn{4}{l}{$^{b}$ {\footnotesize Cumulative number counts.}}\Tstrut\\
\end{tabular}
\label{table:counts}
\end{center}
\end{table}

We then compare our data-points with the most comprehensive literature studies for blank fields at both 450 and 850~$\mu$m 
(\citealt{TC2013b,Casey2013,Geach2013,Wang2017,Zavala2017,Geach2017}).
In Figure~\ref{true_counts} we plot the fit -- Schechter (Sc.) or broken power-law (bpl)\footnote{If a work presented both functions for their fits, 
we selected their Schechter fit. Our results do not depend on this choice.} -- from those works.
Our 450~$\mu$m data well agree with these literature curves\footnote{We remind the reader that -- as already noted in \citet{Casey2013} -- 
the equation (1) of \citet{Geach2013} should be written as $dN/dS=(N'/S')(S/S')^{1-\alpha}\exp(-S/S')$, and the best-fitting parameter $N'$ 
for this 450~$\mu$m data should be quoted as $N'=4900\pm1040$~deg$^{-2}$~mJy$^{-1}$ rather than $N'=490\pm1040$~deg$^{-2}$~mJy$^{-1}$.}, 
while the 850~$\mu$m data-points are above these 
current expectations for blank fields. Especially the more robust data at about 5 and 7~mJy are clearly 
suggesting the presence of higher number counts with respect to the literature values.

To quantify this overdensity of counts at 850~$\mu$m, we 
fit our corrected differential number counts with the functions from each of the literature works allowing only the normalization to vary.
The difference in counts is then estimated through the ratio between the normalizations.
Specifically, 
\begin{itemize}
\item \citet{TC2013b} quoted a best fit with a broken power-law function of the form shown in eqn.~\ref{eqn:diff_counts}, 
with\footnote{For all the fits in the literature, we report only the errors on the parameter $N_0$.} $N_0=120^{+65}_{-45}$~mJy$^{-1}$~deg$^{-2}$, $S_0=6.21$~mJy, $\alpha=2.27$, $\beta=3.71$;
\item \citet{Casey2013} and \citet{Geach2017} reported a Schechter function of the form
\begin{equation}
\label{eqn:Schechter1}
  \frac{dN}{dS} = \frac{N_0}{S_0}\left(\frac{S}{S_0}\right)^{\gamma}{\rm exp}\left(-\frac{S}{S_0}\right), 
\end{equation} 
with $N_0=(3.3\pm1.4)\times10^3$~deg$^{-2}$, $S_0=3.7$~mJy, $\gamma=1.4$, and\footnote{\citet{Geach2017} only showed a Schechter fit to their data in their Figure~15. Here we
thus report the values for a Schechter fit to their data.} $N_0=4550\pm546$~deg$^{-2}$, $S_0=3.40\pm0.21$~mJy, $\gamma=1.97\pm0.08$, respectively;

\item \citet{Zavala2017} preferred a Schechter function of the form
\begin{equation}
\label{eqn:Schechter1}
  \frac{dN}{dS} = \frac{N_0}{S_0}\left(\frac{S}{S_0}\right)^{1-\gamma}{\rm exp}\left(-\frac{S}{S_0}\right),
\end{equation} 
with $N_0=8300\pm300$~deg$^{-2}$, $S_0=2.3$~mJy, $\gamma=2.6$ for all their data.
\end{itemize}

\begin{figure}
\centering
\includegraphics[width=0.95\columnwidth]{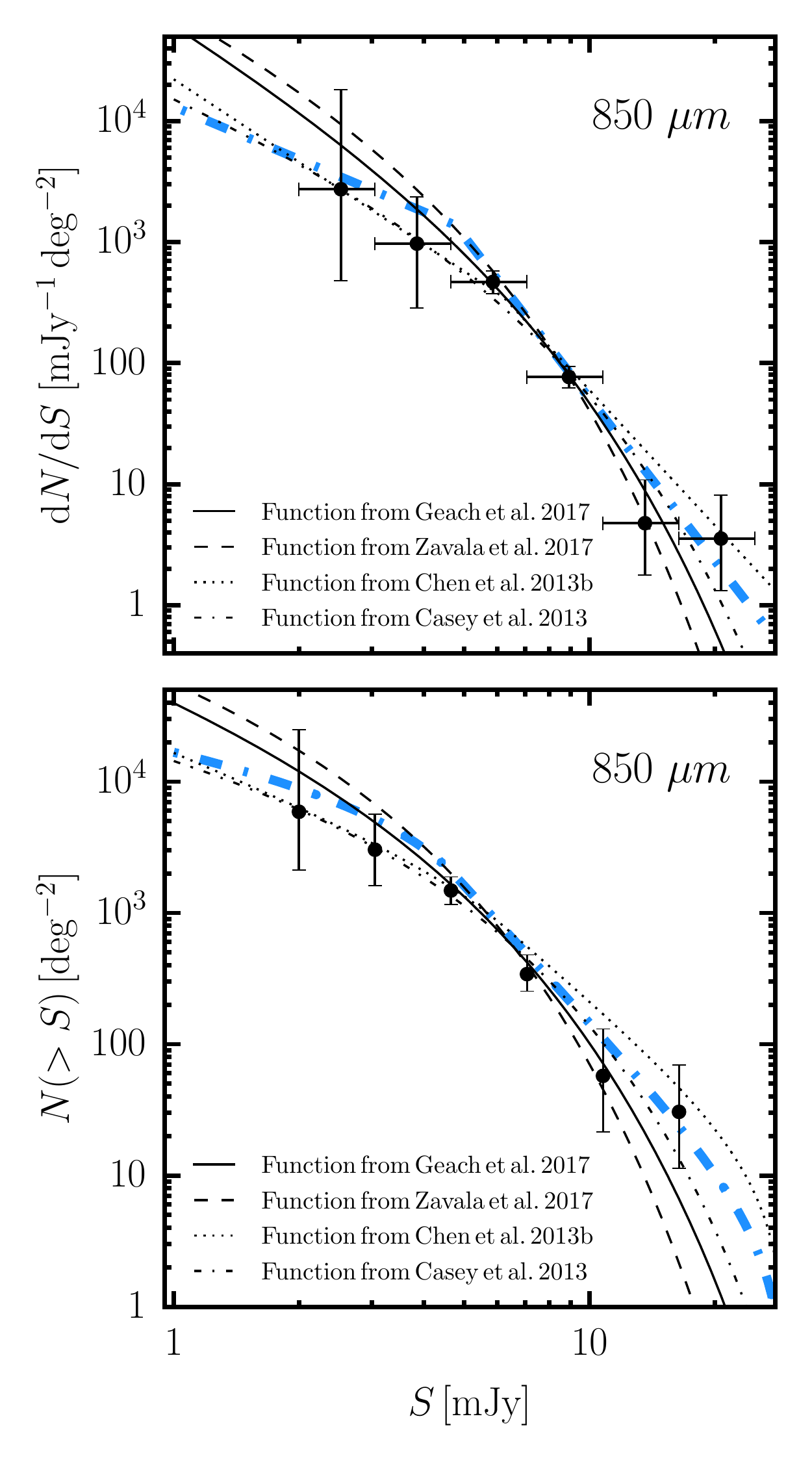}
\caption{Top: fit of the SCUBA-2 MAMMOTH-1 true differential number counts at 850~$\mu$m using the functions given in literature
works for blank fields (\citealt{TC2013b,Casey2013,Zavala2017,Geach2017}) with $N_0$ free to vary. 
Bottom: the SCUBA-2 MAMMOTH-1 true cumulative number counts at 850~$\mu$m compared to the fit models obtained in the top panel.
In both panels, the blue dot-dashed curve is our true number counts curve from Table~\ref{MC_input}. 
All the literature models need a significant increase of their normalization parameter $N_0$ to fit
our data at $850~\mu$m, revealing that the covered effective area is overdense with respect to blank fields. 
We list the values in Table~\ref{table:overdensity}.}
\label{fit_with_Lit}
\end{figure}

We show the results of the fit with free normalizations $N_0$ in Figure~\ref{fit_with_Lit}, and we list in Table~\ref{table:overdensity} the ratio
between the derived normalizations needed to match our data and the literature values. 
From these ratios it is clear that the probed effective area is indeed overdense with respect to blank fields.
On average, around the ELAN MAMMOTH-1 there are $4.0\pm1.3$ times more counts than in blank fields.
In this mean estimate we do \emph{not} include the ratio with respect to \citet{Zavala2017} because this work does not
cover effectively the sources bright-end (their last bin is at 4.9~mJy), probably biasing their fit. We however report the comparison with 
this work for completeness.

\begin{table}
\begin{center}
\caption{Overdensity estimates at 850~$\mu$m from the fit to our true
differential number counts with
literature functions for blank fields.}
\begin{tabular}{ccc}
\hline
\hline
Blank-field function & $N_0^{\rm fit}$ & ratio \Tstrut\Bstrut\\
\hline
\citet{TC2013b}  &    $350\pm30$	  &   $2.9\pm1.4$	\Tstrut \\
\citet{Casey2013}  &  $11800\pm1400$	  &   $3.6\pm1.6$	\Tstrut\\
\citet{Geach2017}  &  $25000\pm3000$	  &   $5.5\pm0.9$	\Tstrut\\
\citet{Zavala2017} &  $75000\pm14000$	  &   $9.1\pm1.8$	\Tstrut\Bstrut \\
\hline
\multicolumn{3}{c}{Mean ratio$^a$~$ = 4.0\pm1.3$}\Tstrut\Bstrut\\
\hline
\multicolumn{3}{c}{Median ratio$^a$~$ = 3.6$}\Tstrut\Bstrut\\
\hline
\hline
\multicolumn{3}{l}{$^a$ {\footnotesize \citet{Zavala2017} is not included due to significantly lower}}\Tstrut\\
\multicolumn{3}{l}{{\footnotesize flux counts probed.}}\Tstrut\\
\end{tabular}
\label{table:overdensity}
\end{center}
\end{table}

\subsection{Position of the catalog sources within the LAE overdensity}
\label{Pos}

\begin{figure*}
\centering
\includegraphics[width=0.95\textwidth]{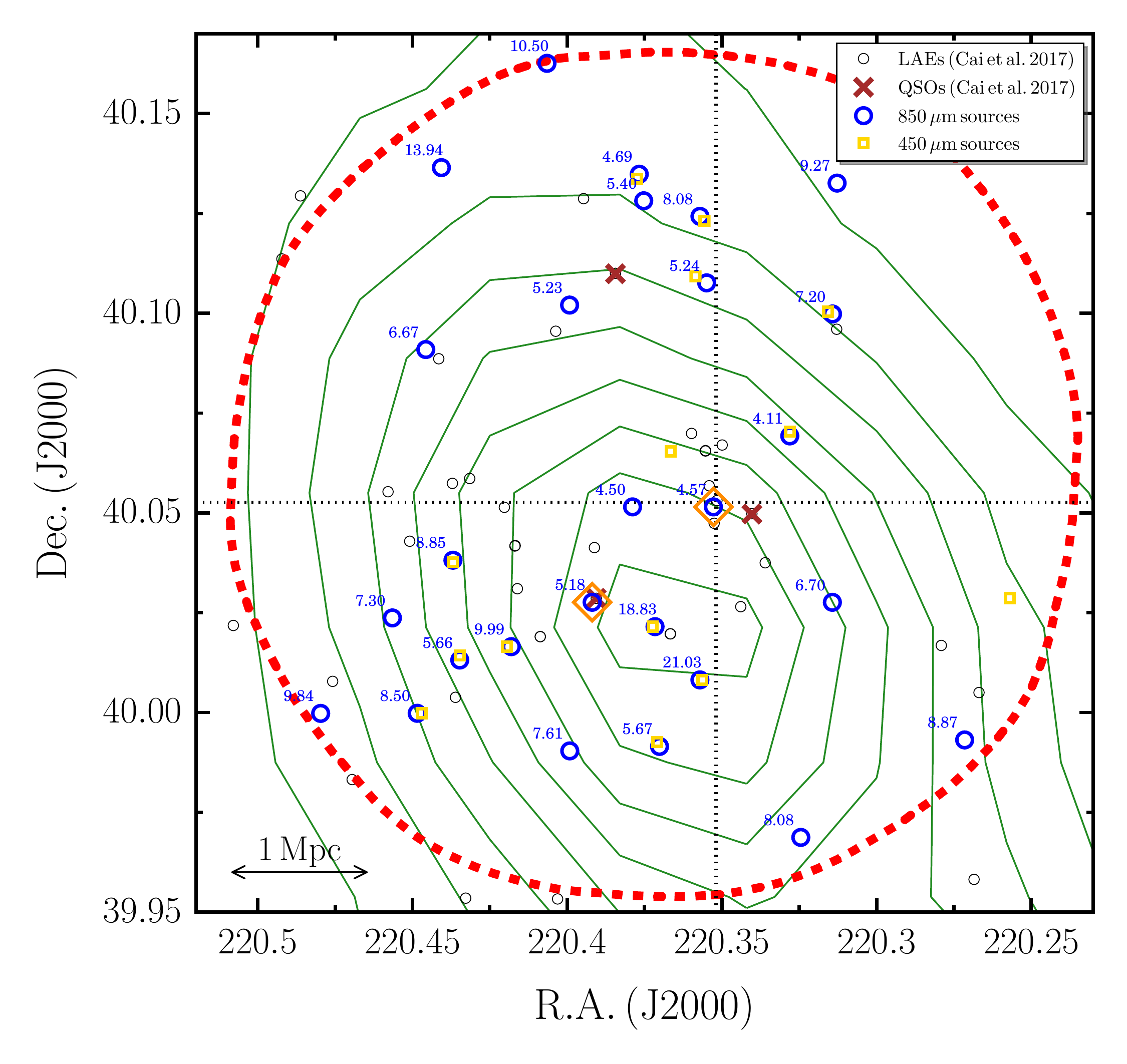}
\caption{Comparison of the location of the known LAEs within the galaxy overdensity BOSS1441 and the SCUBA-2 submillimeter detections in our field-of-view (red; as in
Fig.~\ref{BOSS1441}). We indicate the position of the LAEs (black circles), QSOs in the redshift range $2.30\leq z<2.34$ (brown crosses), 
detections at 850~$\mu$m (blue circles; the number indicates the observed flux in unit of mJy; the size of circles equals the FWHM), and detection at 450~$\mu$m 
(yellow squares; the size of symbol equals the FWHM).
As in Figure~\ref{BOSS1441} we show the density contours (green) for LAEs in steps of 0.1 
galaxies per arcmin$^2$, with the inner density peak of 1.0 per arcmin$^2$. 
We also highlight the position of the ELAN MAMMOTH-1 (dotted crosshair), and the field-of-view of
our SCUBA-2 observations for $3\times$ the central noise (red dashed contour). Overall, the LAEs and submillimeter sources are not associated. 
Only the ELAN MAMMOTH-1 and an additional LAE can be considered as counterparts of a submillimeter detection, MAM-850.14 and MAM-850.16 respectively (highlighted as orange diamonds).
Intriguingly the brightest submillimeter detections coincide with the peak of the BOSS1441 overdensity.}
\label{Comp_with_LAEs}
\end{figure*}

Even though the association to the BOSS1441 overdensity of the sources listed in the SCUBA-2 catalogs has to be confirmed spectroscopically, we
can still search for LAE counterparts to our submillimeter detections, if any. 
In Figure~\ref{Comp_with_LAEs}, we show the location of the 450 (yellow squares) and 850~$\mu$m (blue circles, with fluxes) 
catalog sources along with (i) the position of known LAEs
(black circles; \citealt{Cai2017a}), (ii) the position of known QSOs at $2.30\leq z<2.34$ (brown crosses; \citealt{Cai2017a}), 
and (iii) the LAEs density contours (green; \citealt{Cai2017a}). 
From this figure it is clear that only 2 sources out of the 27 850~$\mu$m detections could be considered to be possibly associated with 
a LAE from the catalogue of \citet{Cai2017a}. This 2 sources (highlighted in orange) are (i) MAM-850.14 close to 
the ELAN MAMMOTH-1 (see Section~\ref{MAMMOTH-1} for a discussion), and (ii) MAM-850.16 close to a LAE at RA=220.3906 and Dec=40.0286, 
with rest-frame equivalent width $EW_0=25.16\pm0.01\AA$, which is actually a $z\simeq2.3$ QSO.

The other 25 LAEs lay at a separation greater than the 850~$\mu$m beam from any of our detections. 
Given the large offsets, the positional uncertainties presented in Section~\ref{pos_err} are not affecting the lack of association between LAEs 
and our submillimeter detections. 
If future follow-up studies confirm the association of most of the SCUBA-2 sources with the BOSS1441 overdensity, 
the lack of submillimeter flux at the location of LAEs is consistent with the usual finding 
that most of the strongly Ly$\alpha$ emitting galaxies are relatively devoid of dust (e.g., \citealt{Ono2010,Hayes2013,Sobral2018}).

In addition, the brightest detections at 850~$\mu$m, MAM-850.1 and MAM-850.2 ($f_{850}^{\rm Deboosted}=18.3\pm2.8$~mJy 
and $f_{850}^{\rm Deboosted}=16.3\pm2.7$~mJy) 
lay intriguingly  close to the peak of the LAEs overdensity.
Their observed ratios between 450 and 850~$\mu$m suggest that these two bright detections are unlikely to be low redshift sources.
Therefore, they probably are associated with the protocluster given the rare alignment with the peak of the LAEs overdensity.

%--------------------------------------------------------------------
\section{Discussion}
\label{sec:disc}
%--------------------------------------------------------------------

\subsection{BOSS1441: a rich and diverse protocluster}
\label{disc:rich}

In the previous sections we have demonstrated the presence of a $\sim 4$ times higher density fluctuation compared 
to blank fields at 850~$\mu$m. In addition, we found that the brightest of our detections are located
at the peak of the LAEs overdensity.
This unique alignments suggest that most of the SCUBA-2 detections are likely associated with the BOSS1441 overdensity (and the ELAN MAMMOTH-1), 
rather than being intervening. 

To test this, we search the available multiwavelength catalogs and build the spectral energy distributions (SEDs) for all the 
sources in our sample.
We thus look for counterparts in the {\it AllWISE} 
Source Catalog\footnote{\url{http://wise2.ipac.caltech.edu/docs/release/allwise/}} (\citealt{Wright2010}) 
at 3.4, 4.6, 12.1, 22.2~$\mu$m (W1, W2, W3, W4), and
in the Faint Images of the Radio Sky at Twenty-cm (FIRST) Survey at 1.4GHz (\citealt{Becker1994}).
This portion of the sky has not been covered by the {\it Herschel} telescope and thus our SCUBA-2 observations are key in covering the far-infrared portion 
of the SED.
To match the different catalogs, we look for counterparts within a 850~$\mu$m beam, and select the closest source.
We found that 8 of our detections have a counterpart in {\it AllWISE}, while none has been detected in FIRST down to the catalog detection limit at each source position ($\approx0.95$ mJy).
To estimate the likelihood of false match for the WISE counterparts, we use the $p$-value as defined in \citet{Downes1986}

\begin{equation}
p = 1 - \exp(-\pi n \theta^{2}),
\end{equation}
where $n$ is the {\it AllWISE} source density within the effective area, $n\simeq0.00134$~sources/arcsec$^2$, and $\theta$ is the angular separation between 
the {\it AllWISE} source and the SCUBA-2 detection\footnote{As the positional uncertainty is small for the 450~$\mu$m band, for our sources we adopt the coordinates at 450~$\mu$m if available.}.
A value of $p<0.05$ usually makes a counterpart reliable (e.g., \citealt{Ivison2002,TC2016}), while $0.05<p<0.1$ makes it tentative (e.g., \citealt{Chapin2009}).
Of the eight counterparts we found that only four are robust, i.e. MAM-850.8, MAM-850.18, MAM-850.26 and MAM-850.27, while the others are tentative.
However, we have shown in Section~\ref{Pos} that MAM-850.16 is likely associated with a quasar at $z=2.30$. As the quasar is detected by WISE and no other close
WISE detections are present, we consider this match secure.
Further, the lack of available radio and/or high resolution submm data prevents us to perform a robust identification of counterparts in our 
recently obtained LBT/LBC $U$, $V$ and $i$ band images (\citealt{Cai2017a}) and in our UKIRT/WIRCAM $J$, $H$, and $K$ band images (Xu et al. in prep.).
We summarize the sources with multiwavelength detections in Table~\ref{SED_values_ALL} , and display for illustration purposes the SEDs of the five sources with
robust counterparts in {\it AllWISE} in Figure~\ref{SEDs_ALL}. 

We leave a detailed classification of our detections 
to future studies encompassing the whole protocluster extent, and better covering the electromagnetic spectrum. 
However, we used the average SED template for SMGs obtained from 99 sources in the ALESS survey (\citealt{daCunha2015})\footnote{\citet{daCunha2015} provides average SEDs made in bins of redshifts, observed ALMA 870~$\mu$m flux, 
average V-band attenuation $A_V$, total dust luminosity (\url{http://astronomy.swinburne.edu.au/~ecunha/ecunha/SED_Templates.html}).} to compute a rough estimate for the 
far-infrared (FIR) luminosity $L_{\rm FIR}$ for each source assuming they are all SMGs at $z=2.32$ (redshift of BOSS1441).
After normalizing the SED template to our SCUBA-2 observations, we found $L_{\rm FIR}\geq4.8\times10^{12}$~L$_{\odot}$ for each of the sources\footnote{We estimated the 
FIR luminosity $L_{\rm FIR}$ for each source as frequently done by integrating the luminosity in the rest-frame range 8-1000~$\mu$m.}.

Next, we compared the volume density implied by our observations with expectations from the current luminosity function of SMGs.
Assuming the effective area of our observations ($\sim336.9$~Mpc$^2$) and the distance interval spanned in redshift by the protocluster ($z=2.3-2.34$; $\sim34.9$~Mpc$^{-3}$), 
the comoving volume targeted by our observations is about $11800$~Mpc$^3$.
If all (75\%) of the detections at 850$\mu$m belongs to the protocluster, their volume density would then be $2.3\times10^{-3}$~Mpc$^{-3}$ ($1.7\times10^{-3}$~Mpc$^{-3}$).
These values are a factor of $>30$ above the volume density expected from the current luminosity function of SMGs with $L_{\rm FIR}\geq4.8\times10^{12}$~L$_{\odot}$ 
($\sim5\times10^{-5}$~Mpc$^{-3}$; \citealt{Casey2014}). Therefore, BOSS1441 is a potentially SMG-rich volume.

We further noticed that the two brightest sources, MAM-850.1 and MAM-850.2, seem to depart from the templates at 850$\mu$m, showing
higher fluxes than expected. This deviation could be explained by allowing a different (higher) redshift, or most probably by the fact that
single-dish submillimeter sources as bright as MAM-850.1 and MAM-850.2 are usually a blend of 
$\geq2$ SMGs once observed with interferometers (\citealt{Karim2013,Simpson2015}).
This explanation is compelling as the two sources sit at the peak of the overdensity. They could thus be groups of 
interacting galaxies, pinpointing the core of the protocluster.

Extremely compact ($20\arcsec$-$40\arcsec$) protocluster cores made of several ($>10$) starbursting galaxies at $z\sim4$ have 
been recently discovered by \citet{Miller2018} and \citet{Oteo2018}. These structures have a global $L_{\rm FIR}\approx10^{14}$~L$_{\odot}$ (\citealt{Miller2018}). 
The BOSS1441 protocluster might thus have similar, but scaled down central structures.  
We can compare this central portion of the overdensity with other known protoclusters at $z\sim2$ (\citealt{Dannerbauer2014,Casey2015,Kato2016}). 
These studies found that spheres with 1~Mpc radius centered at the protocluster core enclose a star-formation rate density of ${\rm SFRD}\sim1000-1500$~M$_{\odot}$~yr$^{-1}$~Mpc$^{-3}$.
Following the approach in those works, we centered a 1~Mpc sphere at the position of MAM-850.2, which is the closest SMG to the peak of the LAE overdensity.
Within this sphere we potentially found six detections (MAM-850.1,MAM-850.2,MAM-850.12,MAM-850.14,MAM-850.16,MAM-850.21), which add up to a total 
star formation rate of ${\rm SFR}=9100$~M$_{\odot}$~yr$^{-1}$, translating to ${\rm SFRD}\approx2200$~M$_{\odot}$~yr$^{-1}$~Mpc$^{-3}$ after dividing by the sphere volume\footnote{We convert $L_{\rm FIR}$ to SFR using the classical conversion in \citet{kennicutt98}.}. 
As our detections are candidate SMGs, this value represents an upper limit. Subtracting the field average value as done in \citet{Kato2016} for $z=2.3$, and assuming that only 75\% of our detections are within the sphere, we obtained ${\rm SFRD}\approx1200$~M$_{\odot}$~yr$^{-1}$~Mpc$^{-3}$.
We conservatively conclude that ${\rm SFRD}=1200^{+1000}_{-1100}$~M$_{\odot}$~yr$^{-1}$~Mpc$^{-3}$, with the lower limit given by the very unlike case that none of the detections (apart
the ELAN MAMMOTH-1 counterpart) are within the sphere. This SFRD value is in agreement with values for protoclusters in the literature (e.g. see Fig.~5 in \citealt{Kato2016}). 

Spectroscopic and interferometric follow-ups are needed to confirm the redshift of our sources, and 
to unveil their nature.
Our analysis however suggests that BOSS1441 is a rich protocluster hosting several LAEs, one ELAN, and likely several SMGs.

\subsection{The Counterpart of the ELAN MAMMOTH-1}
\label{MAMMOTH-1}

\citet{Cai2016} reported the presence of a continuum source at $z=2.319\pm0.004$, named source B, and of a $z=0.16$ AGN, both at the location of the peak of the ELAN MAMMOTH-1.
Source B has been invoked as the powering source of the extended Ly$\alpha$ emission.
At a separation of $4.58\arcsec$ from source B and at $2.75\arcsec$ from the $z=0.16$~AGN, our SCUBA-2 observations resulted in 
a bright detection at 850~$\mu$m, MAM-850.14, with flux density of $f_{850}=4.57\pm0.93$~mJy 
($f_{850}^{Deboosted}=2.83\pm1.03$~mJy), and a $3\sigma$ upper limit of $f_{450}<16.65$~mJy at 450~$\mu$m.
The non-detection at 450~$\mu$m suggests that the emission at 850~$\mu$m is associated with the $z=2.319$ source. 
Indeed, a $z=0.16$ AGN with such a detection at 850~$\mu$m should have a much brighter dust thermal emission at 450~$\mu$m. Specifically, 
if we assume a modified black-body for
optically thin thermal dust emission, a dust temperature of $T_{\rm dust}=45$~K, and the largely used emissivity index $\beta=1.5$ (e.g., \citealt{Casey2012}),
we find that a $z=0.16$ AGN should have $f_{450}=35.3$~mJy (or $21.8$~mJy for the deboosted flux). These fluxes would be detected at high significance 
($\gtrsim4\sigma$) even in our shallow 450~$\mu$m data at the location of the ELAN MAMMOTH-1.

\begin{figure*}
\centering
\includegraphics[width=1.0\textwidth]{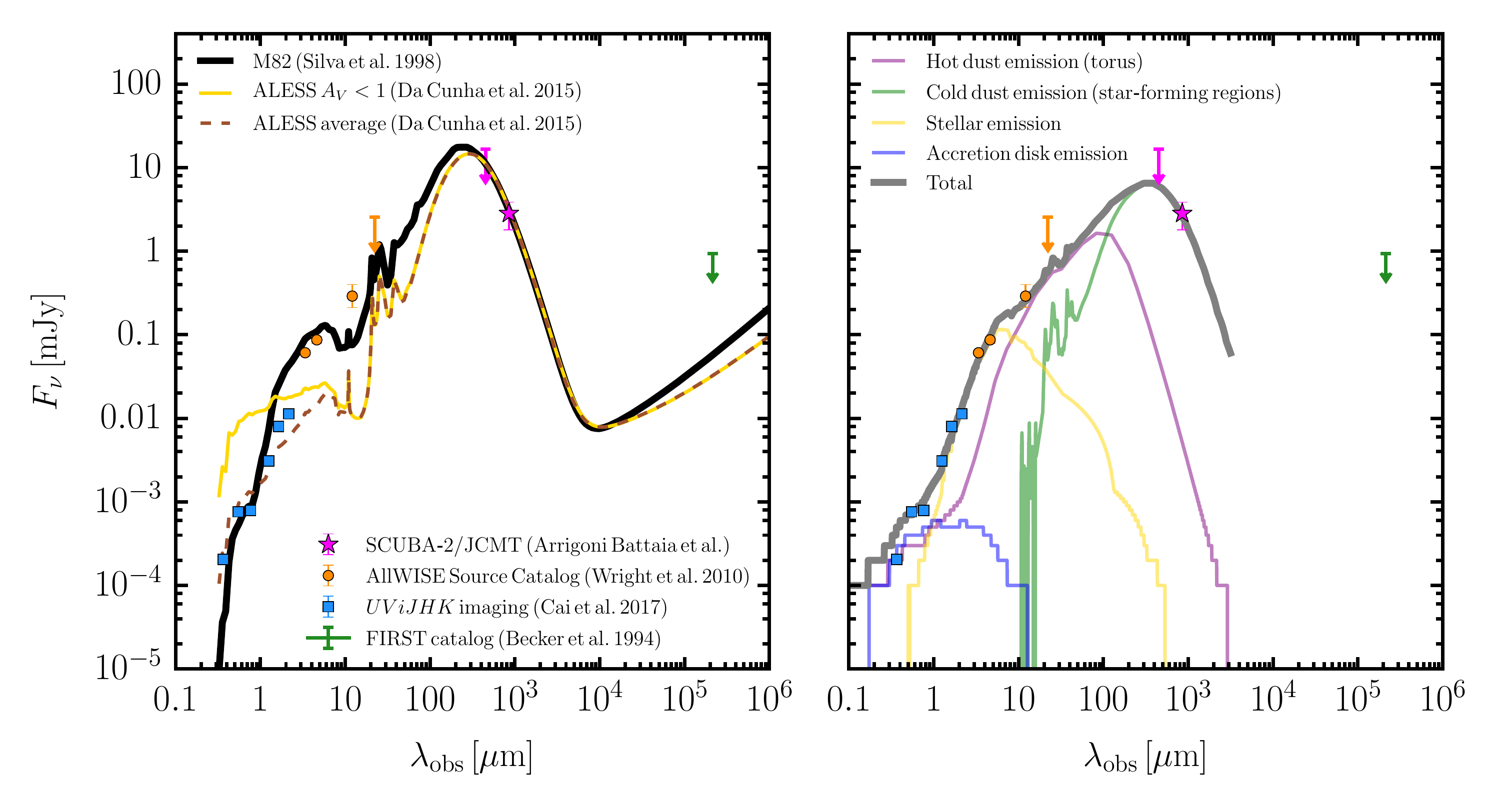}
\caption{SED for the source powering the ELAN MAMMOTH-1 at $z=2.319$ (source B in \citealt{Cai2016}). The data-points are from 
LBT/LBC imaging and UKIRT/WIRCAM (\citealt{Cai2016} and Xu et al. in prep.; blue), AllWISE source catalog (\citealt{Wright2010}; orange), our SCUBA-2 observations (magenta), and the 
FIRST survey (\citealt{Becker1994};green). Left panel: the data-points in comparison to the SED of M82 (\citealt{Silva1998}; black line), of the average ALESS SMGs (dashed brown line), 
and of the average ALESS SMGs with $A_V<1$ (\citealt{daCunha2015}; yellow), 
best fitting our data-points for $z=2.319$. The best agreement is found for the M82 template, but the W3 data-point is underestimated.
Right panel: the data-points in comparison to the best-fit of {\it AGNfitter} (gray line), together with the individual components: hot dust emission (purple), cold dust emission
(green), stellar emission (yellow), and the emission from the accretion disk (blue). The hot dust contribution allows a better fit to the data.}
\label{SEDsourceB}
\end{figure*}

To constrain the nature of source B, we compiled all its data available from the literature, and compared them to known SED.
We summarize all the available observations in Table~\ref{SED_values} \footnote{The $U$, $V$, $i$-band photometry here reported is slightly (within errors) different from 
\citet{Cai2016} because of the image degradation applied to the data to match the UKIRT observations (Xu et al. in prep.).}, while we plot them in Figure~\ref{SEDsourceB}.

To compare these data-points to known SEDs, we fixed the redshift of the source to the redshift $z=2.319$ determined from the \ion{He}{ii} line emission
(\citealt{Cai2016}), and we fitted the data leaving the normalization free.
We first take in consideration the average SED for SMGs obtained from the 99 sources in ALESS (\citealt{daCunha2015}), 
and all the available average SEDs from that publication.
The left panel of Figure~\ref{SEDsourceB} shows this test, highlighting the shortage of emission at the WISE bands for these SED templates (we plot only two to avoid confusion) in comparison to the source B's data-points.
None of the average templates in \citet{daCunha2015} match the W1,W2,W3 data-points, with the SED with $A_V<1$ (yellow) giving the closer values, 
though differing still significantly. 
We then follow the same procedure with the template SED of the local starburst galaxy M82 (\citealt{Silva1998}; solid black line in Figure~\ref{SEDsourceB}).
This template match significantly better the observations, with only the W3 data-point underestimated. Most likely a hotter dust component
powered by an AGN (e.g., \citealt{Silva2004,Fritz2006}) would allow a better match of the data of source B. Indeed \citet{Cai2016} demonstrate that only hard-ioninzing 
sources -- most likely an AGN or a wind -- could power the \ion{He}{ii} and \ion{C}{iv} emission in this object.

To test this interpretation further, we used the publicly available SED fitting code, {\it AGNfitter} (\citealt{CalistroRivera2016}), which adopts a fully Bayesian Markov Chain Monte
Carlo method to model the SEDs of galaxies and AGN. {\it AGNfitter} fits simultaneously the sub-mm to UV photometry decomposing the SED into four
physically motivated components: the AGN accretion disk emission (Big Blue Bump), the hot dust emission from the obscuring structure around
the accretion disk (torus), the cold dust emission from star-forming regions and the stellar populations of the host galaxy\footnote{{\it AGNfitter} does not currently cover the radio
portion of the spectrum. This does not affect our results as we do not have stringent limits at the radio wavelengths.}. Details on the
specific models are presented by \citet{CalistroRivera2016} and references therein.
The right panel of Figure~\ref{SEDsourceB} shows the best fit (in gray) produced by {\it AGNfitter}, with each component highlighted by a different color.
The W3 data-point is now well covered by a composition of the AGN-powered hot dust, star emission and starformation-powered cold dust.

This analysis thus suggests that source B is an enshrouded strong starbursting galaxy, likely hosting an obscured AGN. 
Using the output from {\it AGNfitter}, 
we can separate the far-infrared (FIR) luminosity $L_{\rm FIR}$ (rest-frame 8-1000~$\mu$m) due to the AGN and to star-formation (SF).
We find $L_{\rm FIR}^{\rm AGN}=8.0^{+1.2}_{-6.3}\times10^{11}$~L$_{\odot}$ and $L_{\rm FIR}^{\rm SF}=2.4^{+7.4}_{-2.1}\times10^{12}$~L$_{\odot}$, respectively.
Source B thus meets the generally used criteria to define an UltraLuminous InfraRed Galaxy 
(ULIRG; L$_{\rm 8-1000\mu m}>10^{12}$~L$_{\odot}$; e.g., \citealt{SandersMirabel1996}). Following the classical conversion in \citet{kennicutt98} and 
considering only the star-formation powered emission, one would then obtain a star formation rate of SFR~$=400^{+1300}_{-400}$~M$_{\odot}$~yr$^{-1}$. 

In agreement with the observations and analysis in \citet{Cai2016}, our analysis thus suggests a strong similarity between the source B embedded within the ELAN MAMMOTH-1 and the 
ULIRG sample hosting AGN activity in \citet{Harrison2012}. Source B -- with its obscured AGN and starburst -- can thus easily power the surrounding ELAN, and thus the 
outflow resulting in the velocity offset of 700~km~s$^{-1}$ between the two spectral components in Ly$\alpha$, \ion{He}{ii}, and \ion{C}{iv} (\citealt{Cai2016}). 
The very broad [\ion{O}{iii}] emission presented for the targets in \citet{Harrison2012} however extends to lower distances (15~kpc) with respect 
to the rest-frame UV lines seen in the ELAN MAMMOTH-1 ($\gtrsim30$~kpc; \citealt{Cai2016}). As the ELAN MAMMOTH-1 hosts an ULIRG we thus expect
to see broad [\ion{O}{iii}] emission in its central portion down to similar depths.

Finally, {\it AGNfitter} estimated the stellar mass of source B to be log$(M_{\rm star}/{\rm M_{\odot}})=11.4^{+0.3}_{-0.2}$.
By inverting the halo mass $M_{\rm halo}$ - $M_{\rm star}$ relation in \citet{Moster2013}, we derived that  -- if source B is a central galaxy --
the ELAN MAMMOTH-1 is hosted by a very massive halo of log$(M_{\rm halo}/{\rm M_{\odot}})=15.2^{+1.4}_{-1.6}$.
Given the large uncertainties this result has to be confirmed. However, it certainly highlights the peculiarity of the halo
hosting the ELAN MAMMOTH-1, indicating that it sits at the high-mass end of the halo population at this redshift. 
We further note that the stellar mass of source B is intriguingly close to current estimates for the stellar mass of host galaxies of HzRGs
(log$(M_{\rm star}/{\rm M_{\odot}})\simeq 11 - 11.5$; \citealt{Seymour2007,DeBreuck2010}). HzRGs are currently thought to reside in
massive halos of mass log$(M_{\rm halo}/{\rm M_{\odot}})\approx 13$ (e.g., \citealt{Stevens2003}). The halo hosting the ELAN MAMMOTH-1 could thus be similarly massive
or exceed such halos.

\begin{sidewaystable}
\begin{center}
\caption{Data for source B, counterpart of the MAMMOTH-1 ELAN.}
\scalebox{1}{
\scriptsize
\setlength\tabcolsep{4pt}
\begin{tabular}{cccccccccccccccc}
\hline
\hline
 Name     &  $U$	     &   $V$            &  $i$	           &  $J$	      &  $H$		 & $K$		    & W1               & SNR$_{\rm W1}$ &   W2    & SNR$_{\rm W2}$ &	W3   & SNR$_{\rm W3}$ &   W4	& SNR$_{\rm W4}$ &    1.4GHz  \\
          & (AB)             &  (AB)            &  (AB)            &  (AB)	      &  (AB)		 & (AB)		    &(Vega)$^{a}$      &		& (Vega)  &		   &  (Vega) &  	      &  (Vega) &		 &	      \\
          & ($10^{-4}$~mJy)  &  ($10^{-4}$~mJy) &  ($10^{-4}$~mJy) &  ($10^{-4}$~mJy) &  ($10^{-4}$~mJy) & ($10^{-4}$~mJy)  &(mJy)             &		&  (mJy)  &		   &  (mJy)  &  	      &  (mJy)  &		 &    (mJy)   \\

\hline
Source B  & 25.62$\pm$0.10 & 24.20$\pm$0.06 & 24.16$\pm$0.11 & 22.67$\pm$0.10 & 21.65$\pm$0.07 & 21.26$\pm$0.08 & 16.76$\pm$0.07 & 15.6 & 15.73$\pm$0.10 & 10.7 & 12.50$\pm$0.34 & 3.2 & <8.8$^b$ & 1.7 &      \Tstrut \\ 
          &  2.1$\pm$0.2  &  7.6$\pm$0.4    &  7.9$\pm$0.8   & 30.9$\pm$2.8   & 79.7$\pm$5.1   & 113.4$\pm$8.3  & 0.061$\pm$0.004  &      & 0.087$\pm$0.008  &	  & 0.29$\pm$0.09  &     & <2.6$^b$ &     & <0.94$^c$ \Bstrut \\ 
\hline
\hline
\multicolumn{13}{l}{$^a$ We converted the Vega magnitude in the AllWise Source Catalog to mJy following \citealt{Wright2010} and a source with $F_{\nu}\propto\nu^{-2}$ (see their table 2).}\Tstrut\\ 
%\footnote{\url{http://wise2.ipac.caltech.edu/docs/release/allsky/expsup/sec4_4h.html#conv2flux}}.}\Tstrut\\
\multicolumn{13}{l}{$^b$ This value is the 95\% confidence brightness upper limit from the AllWise Source Catalog.}\Tstrut\\
\multicolumn{13}{l}{$^c$ Detection limit of the FIRST catalog at source position.}\Tstrut\\
\end{tabular}
}
\label{SED_values}
\end{center}
\end{sidewaystable}

\section{Summary}
\label{sec:summ}

We are conducting a survey of all the known ELANe (\citealt{cantalupo14,hennawi+15,Cai2016,FAB2018}) with the 
JCMT and APEX telescopes to assess the presence of starburst activity in these systems and their environments.
In this work we focused on the SCUBA-2/JCMT data at 450 and 850~$\mu$m obtained for an effective area of
$\sim127$~arcmin$^2$ around the ELAN MAMMOTH-1 at $z=2.319$ (\citealt{Cai2016}), and thus targeting the known peak area
of the LAE overdensity BOSS1441 (\citealt{Cai2017a}).
Thanks to this dataset we found that

   \begin{enumerate}
      \item the 850~$\mu$m source counts are $4.0\pm1.3$ times higher than in blank fields (\citealt{TC2013b,Casey2013,Geach2017}),
       confirming also in obscured tracers the presence of an overdensity. Intriguingly,
       the two brightest submillimeter detections, MAM-850.1 and MAM-850.2, are located at the peak of the LAE overdensity, possibly
       pinpointing the core of the protocluster and multiple mergers/interactions (e.g., \citealt{Miller2018}). 
       The association of the discovered 
       submillimeter sources with BOSS1441 needs however a spectroscopic confirmation.
      \item the continuum source at the center of the ELAN MAMMOTH-1, source B (\citealt{Cai2016}),
      is associated to a strong detection at 850~$\mu$m, MAM-850.14, with flux density of
      $f_{850}=4.6\pm0.9$~mJy ($f_{850}^{\rm Deboosted}=2.8\pm1.0$~mJy) and a
      $3\sigma$ upper limit of $f_{450}<16.6$~mJy at 450~$\mu$m. Together with
      the data from the literature, the SED of source B agrees with a strongly starbursting galaxy
      hosting an obscured AGN, and having a FIR luminosity of $L_{\rm FIR}^{\rm SF}=2.4^{+7.4}_{-2.1}\times10^{12}$~L$_{\odot}$. 
      Source B is thus an ULIRG with a star-formation rate of SFR~$=400^{+1300}_{-400}$~M$_{\odot}$~yr$^{-1}$ 
      assuming the classical \citet{kennicutt98} calibration.
      Such a source -- containing both 
      an AGN and a violent starburst -- is able to power the hard photoionization plus outflow scenario depicted 
      in \citet{Cai2016}.
   \end{enumerate}

The acquisition of wide-field multiwavelength data (X-ray, UV, optical, submillimeter, radio) 
is key in painting a coherent and detailed picture of a protocluster, and ultimately to 
understand the assembly of massive galaxies within the cosmic nurseries of the soon-to-be large clusters.
The results of this pilot project are encouraging and reflect the importance of such a multi-wavelength approach in 
fully comprehending the ELAN phenomenon and the environment in which they reside.

\begin{acknowledgements}
We thank the referee Yuichi Matsuda for his careful read of the manuscript.
The James Clerk Maxwell Telescope is operated by the East Asian Observatory on behalf 
of The National Astronomical Observatory of Japan; Academia Sinica Institute of Astronomy and Astrophysics; 
the Korea Astronomy and Space Science Institute; the Operation, Maintenance and Upgrading Fund for Astronomical Telescopes 
and Facility Instruments, budgeted from the Ministry of Finance (MOF) of China and administrated by the Chinese Academy of Sciences (CAS), 
as well as the National Key R\&D Program of China (No. 2017YFA0402700). 
Additional funding support is provided by the Science and Technology Facilities Council 
of the United Kingdom and participating universities in the United Kingdom and Canada.
This publication makes use of data products from the Wide-field Infrared Survey Explorer, which is a joint project 
of the University of California, Los Angeles, and the Jet Propulsion Laboratory/California Institute of Technology, and NEOWISE, 
which is a project of the Jet Propulsion Laboratory/California Institute of Technology. 
WISE and NEOWISE are funded by the National Aeronautics and Space Administration.
M.F. acknowledges support by the Science and Technology Facilities Council [grant number ST/P000541/1]. 
This project has received funding from the European Research Council (ERC) under the European Union's Horizon 2020 
research and innovation programme (grant agreement No. 757535). 
I.R.S. acknowledges support from the ERC Advanced Grant {\it DUSTYGAL} (321334) and STFC (ST/P000541/1). 
Y.Y.'s research was supported by Basic Science Research Program through the National Research Foundation of Korea (NRF) 
funded by the Ministry of Science, ICT \& Future Planning (NRF-2016R1C1B2007782).
The authors wish to recognize and acknowledge the very significant cultural
role and reverence that the summit of Mauna Kea has always
had within the indigenous Hawaiian community. We are most
fortunate to have the opportunity to conduct observations from
this mountain.
\end{acknowledgements}

\bibliographystyle{aa} 
\bibliography{allrefs}

\begin{appendix}

\section{{\it AllWISE} counterparts to our SCUBA-2 detections}

\begin{figure*}
\centering
\includegraphics[width=0.9\textwidth]{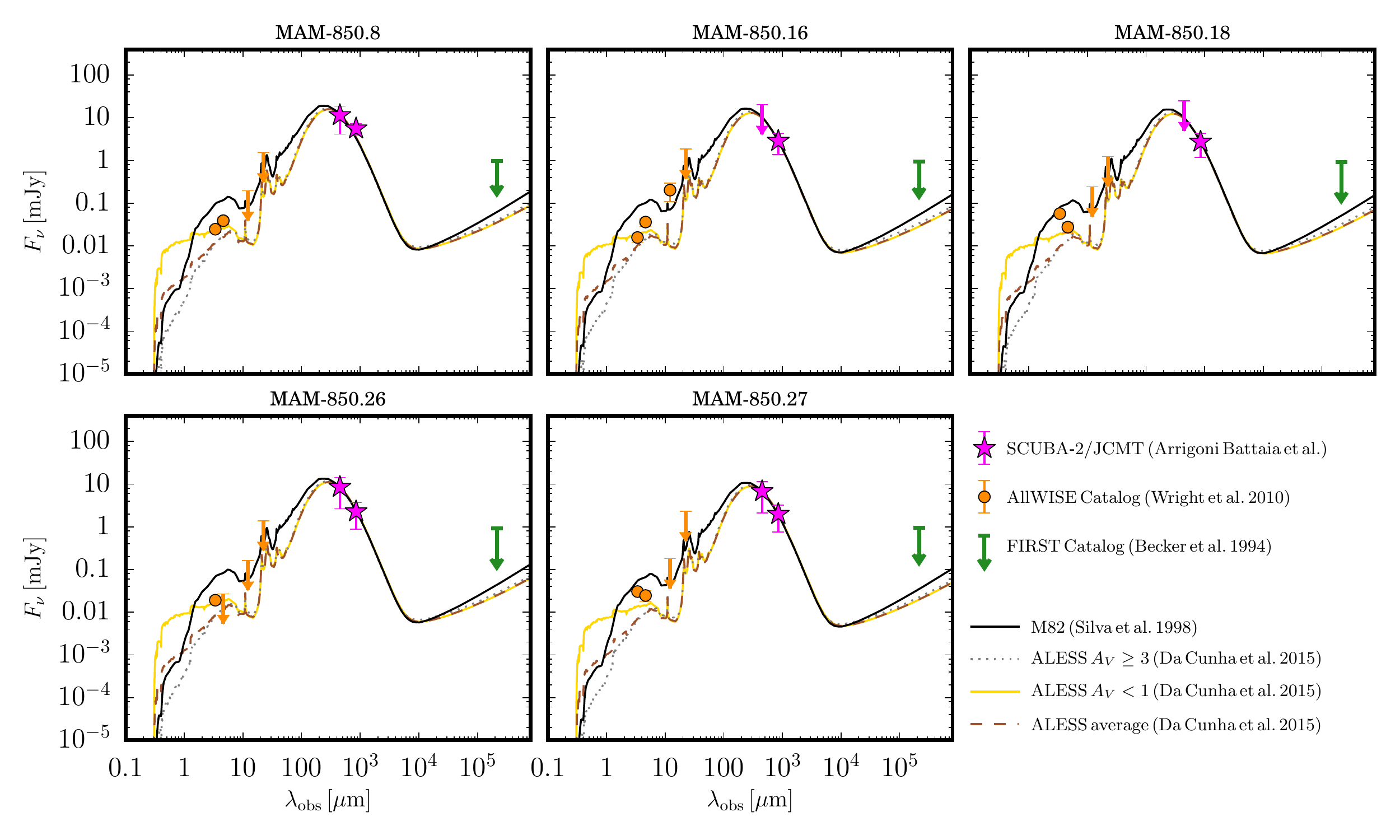}
\caption{SEDs for the SCUBA-2 sources with robust {\it AllWISE} counterparts. The data-points (see Table~\ref{SED_values_ALL}) are 
from AllWISE source catalog (\citealt{Wright2010}; orange), 
our SCUBA-2 observations (magenta), and the 
FIRST survey (\citealt{Becker1994}; green). We show the SED of M82 (\citealt{Silva1998}; black line), of the average ALESS SMG (dashed brown line), 
and the average ALESS SMG with $A_V<1$, and $A_V \geq 3$ (\citealt{daCunha2015}; yellow), 
best fitting only our SCUBA-2 data assuming $z=2.32$. }
\label{SEDs_ALL}
\end{figure*}

In this appendix we show the multiwavelength dataset for {\it AllWISE} counterparts to our SCUBA-2 detections.
Specifically, in Table~\ref{SED_values_ALL} we list (i) the likelihood of false match for each {\it AllWISE} counterpart, i.e. the $p$-value (see Section~\ref{disc:rich}), (ii) the magnitudes from the 
{\it AllWISE} catalog, (iii) the flux from the FIRST survey, and (iv) the rough estimate of $L_{\rm FIR}$ for each source.
The four sources with $p<0.05$ are considered robust, while the others are tentative. We further consider robust the 
match with MAM-850.16 has it is clearly associated with a quasar at $z=2.30$ (see Section~\ref{Pos} for more details).
Finally, for illustration purposes, in Figure~\ref{SEDs_ALL} we show the SED of the five sources with robust {\it AllWISE} counterparts along with
the template SED of M82 (\citealt{Silva1998}; black line), of the average ALESS SMGs (dashed brown line), 
and the average ALESS SMGs with $A_V<1$, and $A_V \geq 3$ (\citealt{daCunha2015}; yellow). All the template SEDs have been
normalized to the SCUBA-2 data assuming $z=2.32$. We will perform a more detailed analysis of the SED in future works encompassing the 
full extent of the protocluster, and covering a broader range of the electromagnetic spectrum.

%\begin{sidewaystable*}
\begin{table*}%[h]
\begin{center}
\caption{Data for sources SEDs presented in Fig.~\ref{SEDs_ALL}.}
\scalebox{1}{
\scriptsize
\setlength\tabcolsep{4pt}
\begin{tabular}{lccccccccccc}
\hline
\hline
 Name 		&    $p^{a}$     &  W1       & SNR$_{\rm W1}$ &   W2	& SNR$_{\rm W2}$ &    W3   & SNR$_{\rm W3}$ &	W4    & SNR$_{\rm W4}$ &    1.4GHz &  $L_{\rm FIR}^{e}$\\
      		&            & (Vega)$^{b}$&		    &	(Vega)  &		 &  (Vega) &		    &  (Vega) & 	       &   (mJy)   & ($10^{12}$~L$_{\odot}$) \\
\hline
MAM-850.7 	&   0.052    &  16.955$\pm$0.086	 &   12.6	 &   17.071$\pm$0.300	 &   3.6 &   <12.839$^c$	 &   0.2 &   <8.853$^c$  &  1.3  &  <0.94$^d$ & 15.4	\\
MAM-850.8 	&   0.017    &  17.746$\pm$0.162	 &    6.7	 &   16.608$\pm$0.207	 &   5.3 &   <12.933$^c$	 &   0.0 &   <9.315$^c$  &  0.2  &  <0.96$^d$ & 8.68	\\
MAM-850.10	&   0.055    &  17.502$\pm$0.134	 &    8.1	 &   16.008$\pm$0.127	 &   8.6 &   <13.069$^c$	 &   -1.2&   <9.379$^c$  &  0.2  &  <0.94$^d$ & 7.89	\\
MAM-850.16	&   0.091    &  18.232$\pm$0.244    &	 4.4	    &	16.691$\pm$0.210    &	5.2 &	12.898$\pm$0.495    &	2.2 &	<9.12$^c$   &  0.4  &  <0.94$^d$ & 7.04    \\
MAM-850.18	&   0.041    &  16.834$\pm$0.073    &	15.0	    &	16.988$\pm$0.287    &	3.8 &	<12.704$^c$	    &	0.5 &	<9.565$^c$  & -0.5  &  <0.91$^d$ & 6.75    \\
MAM-850.26	&   0.012    &  18.017$\pm$0.202    &	5.4	    &	<17.008$^c$	    &	2.0 &	<13.130$^c$	    &  -0.6 &	<9.437$^c$  &  -1.1 &  <0.91$^d$ & 6.05    \\
MAM-850.27	&   0.014    &  17.51$\pm$0.134     &	8.1	    &	17.106$\pm$0.309    &	3.5 &	<13.017$^c$	    &  -0.5 &	<8.877$^c$  &  1.5  &  <0.94$^d$ & 4.83  \Bstrut   \\
\hline
\hline
\multicolumn{12}{l}{$^a$ $p$-value estimated following \citet{Downes1986}.}\Tstrut\\ 
\multicolumn{12}{l}{$^b$ We converted the Vega magnitude in the AllWise Source Catalog to mJy following \citealt{Wright2010} and a source with $F_{\nu}\propto\nu^{-2}$ (see their table 2).}\Tstrut\\ 
%\footnote{\url{http://wise2.ipac.caltech.edu/docs/release/allsky/expsup/sec4_4h.html#conv2flux}}.}\Tstrut\\
\multicolumn{12}{l}{$^c$ This value is the 95\% confidence brightness upper limit from the AllWise Source Catalog.}\Tstrut\\
\multicolumn{12}{l}{$^d$ Detection limit of the FIRST catalog at source position.}\Tstrut\\
\multicolumn{12}{l}{$^e$ Far-infrared luminosity measured by integrating the fitted average ALESS SED (Fig.~\ref{SEDs_ALL}) in the rest-frame range 8-1000~$\mu$m.}\Tstrut\\
%\multicolumn{12}{l}{$^e$ $3\sigma$ upper limit.}\Tstrut\\
\end{tabular}
}
\label{SED_values_ALL}
\end{center}
\end{table*}
%\end{sidewaystable*}

\end{appendix}

\end{document}